# Surface-Enhanced Raman Scattering from Au Nanorods, Nanotriangles, and Nanostars with Tuned Plasmon Resonances

Boris N. Khlebtsov,[1] Andrey M. Burov,[1] Sergey V. Zarkov,[1] Nikolai G. Khlebtsov,[1,2,*]

[1]Institute of Biochemistry and Physiology of Plants and Microorganisms, "Saratov Scientific Centre of the Russian Academy of Sciences," 13 Prospekt Entuziastov, Saratov 410049, Russia

[2]Saratov State University, 83 Ulitsa Astrakhanskaya, Saratov 410012, Russia

[*]To whom correspondence should be addressed. E-mail: (NGK) khlebtsov@ibppm.ru

**Abstract.** Electromagnetic theory predicts that the optimal value of the localized plasmon resonance (LPR) wavelength for the maximal SERS enhancement factor (EF) is half the sum of the laser and Raman wavelengths. For small Raman shifts, the theoretical EF scales as the fourth power of the local field. However, experimental data often disagree with these theoretical conclusions, leaving the question of choosing the optimal plasmon resonance for the maximal SERS signal unresolved. Here, we present experimental data for gold nanorods (AuNRs), gold nanotriangles (AuNTs), and gold nanostars (AuNSTs) simulating 1D, 2D, and 3D plasmonic nanostructures, respectively. The LPR wavelengths were tuned by chemical etching within 550-1050 nm at constant number concentrations of the particles. The particles were functionalized with Cy7.5 and NBT, and the dependence of the intensity at 940 cm$^{-1}$ (Cy7.5) and 1343 cm$^{-1}$ (NBT) on the LPR wavelength was examined for laser wavelengths of 633 nm and 785 nm. The electromagnetic SERS EFs were calculated by averaging the product of the local field intensities at the laser and Raman wavelengths over the particle surface and their random orientations. The calculated SERS plasmonic profiles were redshifted compared to the laser wavelength. For 785-nm excitation, the calculated EFs were five to seven times higher than those for 633-nm excitation. With AuNR@Cy7.5 and AuNT@ Cy7.5, the experimental SERS was 35-fold stronger than it was with NBT-functionalized particles, but with AuNST@Cy7.5 and AuNST@NBT, the SERS responses were similar. With all nanoparticles tested, the SERS plasmonic profiles after 785 nm excitation were slightly blue-shifted, as compared with the laser wavelength, possibly owing to the inner filter effect. After 633-nm excitation, the SERS profiles were redshifted, in agreement with EM theory. In all cases, the plasmonic EF profiles were much broadened compared to the calculated ones and did not follow the four-power law.

**Introduction.**

Strong local field enhancement near plasmonic nanoparticles is the primary physical mechanism behind the surface-enhanced Raman scattering (SERS).[1,2] Recent advances in wet-chemical technologies[3] provide great possibilities for precise tuning of the localized plasmon resonance (LPR) to any desired wavelength from blue (~400 nm) to NIR.[4] These synthesis advances raise the following important question:[5] Which excitation wavelength maximizes the SERS response for a given LPR nanostructure? There are two possible approaches to answer. The first one is the so-called plasmon-sampled SER excitation spectroscopy (PS-SERES) in which a set of substrates such as nanosphere-lithography (NL) Ag prisms,[6] Au nanostar (AuNST)[7] or Au nanorod (AuNR) colloids[8] with different LPRs is excited by a fixed laser wavelength to find the sample with the maximal SERS response. The second approach is the wavelength-scanned SER excitation spectroscopy (WS-SERES) in which light with variable wavelengths excites a particular Ag-island film,[9, 10] Ag-NL,[11] or vertically aligned AuNRs[12] with a specific LPR spectral profile. This method seems straightforward but is more technically challenging because of the limited laser source with variable wavelength. Nevertheless, the multiple-laser-wavelength excitation of AuNRs functionalized with different dyes allows for SERS bioimaging of cells within the 514-1064 nm spectral band.[13]

Although it is usually believed that excitation at the LPR wavelength gives maximal SERS response,[14,15] there is experimental evidence for optimal excitation wavelengths that are slightly blue shifted, as compared with the LPR.[16] What is more, Weitz et al.[9] proposed a phenomenological relation to approximate the relative SERS intensity, from which the optimal LPR resonance wavelength was found[17] to be half of the excitation and Raman scattering wavelengths $\lambda_{LPR} = (\lambda_{ex} + \lambda_{RS})/2$ (the absorption spectrum is assumed to have the Lorentzian shape). This rule has been confirmed in experiments with Ag island film substrates and, more precisely, with Ag-NL substrates in PS-SERES,[6] WS-SERES,[11] and WS-SERRES[18] experiments.

According to theoretical treatment by Kerker et al.,[19] SERS is a twofold process involving (1) LPR enhancement of the incident light at the excitation wavelength, $\lambda_{ex}$, and (2) enhancement of the dipole Raman scattering by the same plasmonic particle at the Stokes-shifted Raman wavelength, $\lambda_{RS}$. This physical picture has been confirmed recently by Kumar et al.[20] For the twofold SERS process, the enhancement factor EF is proportional to the squared local field amplitudes taken at the incident and Raman scattering wavelengths $EF \sim |E_{loc}(\lambda_{ex})|^2 |E_{loc}(\lambda_{RS})|^2$. By neglecting the Stokes shift, we arrive at the well-known four-power law[2]: $EF \sim |E_{loc}(\lambda_{ex})|^4$.

Simple FDTD calculations for AuNRs[8] predict a strong decrease in $EF \sim |E_{loc}(\lambda_{ex})|^4$ factor when the excitation wavelength is moderately detuned from the LPR wavelength. However, the existing experimental data do not support this prediction. In particular, accurate measurements and electromagnetic simulations of SERS EFs for precisely controlled AuNR ensembles[8] showed almost constant values of EFs for variable LPRs between 700 and 900 nm at a fixed laser wavelength $\lambda_{ex} = 785$ nm. One could assume that such a weak dependence of the EF on LPR[8] is due to ensemble averaging over the particle parameters. However, similar weak LPR dependence can be derived from single-particle AuNR SERS spectra reported by Lin et al.[21, 22] In the same line, Doherty et al.[12] reported weak excitation-wavelength variations of EF within 650-800 nm for vertically aligned and strongly interacting AuNRs. In all the above cases, there was little correlation between nonlinear SERS response and linear extinction spectra.

Thus, the simple four-power law can be violated in actual experiments for different reasons, including dominated hot-spot contributions for particle-coupled substrates and due to orientation, surface, and shape averaging for colloidal systems. Therefore, further studies on the relation between excitation wavelength and LPR for SERS substrates with precisely controlled geometrical and plasmonic parameters are strongly desirable.

Here, we discuss a PS-SERES study with well-defined experimental models obtained by controllable etching of AuNRs, AuNTs, and AuNSTs. In a sense, these particles exemplify 1D, 2D, and 3D plasmonic structures, respectively. Using the chemical etching method[23, 24] we fabricated colloids with equal particle concentrations and variable LPR wavelengths from 1020 to 580 nm. By contrast to our previous reexamination study with etched AuNRs of 26 nm in diameter,[8] here we prepared more thin AuNRs with an average diameter of 13.8 nm. They were functionalized with 1,4-nitrobenzene thiol (NBT) and Cy-7.5 molecules, and SERS spectra were measured under 633 nm and 785 nm laser excitation. In particular, we show that the SERS response from Cy-7.5-functionalized nanoparticles is 35-fold stronger than that for NBT-functionalized ones. Our main finding is that all 1D-3D colloidal SERS substrates demonstrate maximal response near LPRs close to the excitation wavelengths 633 and 785 nm, but the EF variations are much less than those expected from the electromagnetic simulations.

**Experimental and Theoretical Methods**

**Materials.** The following reagents were used without additional purification: Cetyltrimethylammonium bromide (CTAB, > 98.0%), cetyltrimethylammonium chloride (CTAC, 25% water solution), L-ascorbic acid (AA, >99,9), hydrochloric acid (HCl, 37 wt.% in water), thiolated polyethylene glycol (mPEG-SH, 99%), polyvinylpyrrolidone (PVP, $M_w$=40,000), hydroquinone (HQ, 99%), 1,4-nitrothoiphenol (technical grade), sodium iodide (99.9%) and sodium borohydride (NaBH$_4$, 99%) were purchased from Sigma-Aldrich. Cy7.5–amine was obtained from Lumiprobe. Hydrogen tetrachloroaurate trihydrate (HAuCl$_4$·3H$_2$O) and silver nitrate (AgNO$_3$, >99%) were purchased from Alfa Aesar. Ultrapure water obtained from a Milli-Q Integral 5 system was used in all experiments.

**Synthesis and etching of gold nanorods.** Gold nanorods were obtained by a modified protocol,[25] which produces thin, long rods. Au seeds were obtained by adding aqueous sodium borohydride (10 mM, 0.6 mL) to an aqueous solution containing CTAB (0.1 M, 10 mL) and HAuCl$_4$ (10 mM, 0.25 mL). Silver nitrate (3.5 mL, 0.1 M) was added to a solution of HAuCl$_4$ (500 mL, 0.5 mM) in 0.1 M CTAB, followed by the addition of an aqueous solution of hydroquinone (25 mL, 0.1 M). The resulting mixture was stirred by hand until it became clear. Then, 8 mL of seeds were added, mixed, and left overnight at 30° C without stirring. A significant drawback of protocol[25] is a large percentage (10–15%) of impurity particles. Here, we developed a modified version with the purification step in which the colloid was centrifuged at 10,000 g for 20 minutes, and the residue was redissolved in 30 mL of 200 mM CTAC and left without stirring for 1 hour. The nanorods were aggregated on the bottom and walls of the test tube in the form of a brown film. The supernatant containing particles of other shapes was removed. The nanorods were redissolved in 50 mM CTAB to a concentration corresponding to the LPR optical density of 2 measured in a 2 mm cuvette. To completely remove the impurity particles, the purification procedure was repeated twice. TEM analysis of the samples (Figures S1 A, B) showed the presence of thin nanorods with an average thickness of 13.8 nm and a length of about 70 nm. A distinctive feature of our protocol is almost zero percentage of non-target particles (0.5%) and the high-quality factor of the extinction spectrum in terms of the ratio of extinctions at the longitudinal and transverse LPR wavelengths (> 6).

To obtain AuNR samples with different LPR wavelengths, 50 mL of AuNR colloid in 50 mM CTAB was titrated with 2 mM HAuCl$_4$ solution (by adding 20 μL solution every 10 minutes).

Before each addition, the extinction spectrum was measured, and if the desired LPR wavelength was reached, 3 mL of the sample was taken from the colloid. A total of 13 samples were obtained with LPR wavelength from 1017 nm to 580 nm. After synthesis was completed, 3 mL of each selected sample was divided into two tubes of 1.5 mL.

**Functionalization of gold nanorods.** To functionalize AuNR with NBT, 10 μL of 2 mM ethanol solution of NBT was added to 1.5 mL of nanorods. The mixture was incubated for 30 min. Then, the nanorods were centrifuged at 10,000 rpm for 5 min and resuspended in 2 mL of 10 mM CTAC. To functionalize Au nanorods with Cy7.5-amine (from now on, Cy7.5), they were first functionalized with thiolated PEG. For this purpose, 20 μL of 1 mM m-PEG-thiol was added to 1.5 mL of nanorods in 50 mM acetate buffer. The mixture was incubated at 37 °C for 2 hours. Then, the nanorods were centrifuged at 10,000 rpm for 5 min and resuspended in water. Ten μL of Cy7.5 solution in ethanol (50 μg/mL) was added to 1.5 mL nanorod solution, incubated overnight, then centrifuged at 10,000 rpm for 5 min and resuspended in water. Checking the supernatant for extinction showed the absence of free Cy 7.5 molecules. Thus, we assumed all Cy7.5 molecules were adsorbed on the nanorod surface.

**Synthesis, etching, and functionalization of gold nanotriangles.** Gold nanotriangles (AuNTs) were obtained according to an adapted protocol initially proposed by Mirkin and co-workers[26, 27] (see also several recent modifications and improvements).[28-32] First, a 0.05 M CTAB and 50 μM NaI mixture was prepared. Next, gold nuclei 5 nm in diameter were obtained. For this, 0.25 mL of 0.01 M $HAuCl_4$, 0.5 mL of 0.01 M sodium citrate, and 0.3 mL of 0.01 M $NaBH_4$ were added to 20 mL of water with vigorous stirring for 1 min at 1200 rpm. Then, 25 mL of 0.01 M $HAuCl_4$, 5 mL of 0.1 M NaOH, 5 mL of 0.1 M ascorbic acid, and 2 mL of seed particles were added to 1 L of the prepared CTAB/NaI solution. Besides AuNTs, the resulting colloid contained about 90% particles of other shapes. Therefore, one needs a purification step, thus explaining the sizeable 1L volume of the initial preparation. To remove non-target particles, we use a procedure based on the different colloidal stability of particles of different shapes in saline solutions. 3 mL of 2 M NaCl was added to 1 L of AuNTs colloid, which was left without stirring for 1 day. As a result, AuNTs aggregated and deposited on the bottom as a green film. The supernatant containing particles of other shapes was removed. AuNTs were redissolved in 50 mM CTAB to a concentration corresponding to the LPR optical density 2 in a 2 mm cuvette. The purification procedure was carried out 1 time, avoiding significant losses of the target product. The functionalization and etching protocol for AuNTs was similar to that for nanorods. A total of 8 samples of AuNT particles were obtained.

**Synthesis, etching, and functionalization of gold nanostars**. AuNSTs were synthesized by the seed-mediated two-stage protocol as described in Ref.[7] with minor modifications. At the first stage, gold nanospheres with an average diameter of 15 nm were synthesized using the citrate reduction method described by Frens,[33] which were used as seeds for synthesizing gold nanostars. To do this, 238 mL of water boiled in the Erlenmeyer flask while stirring on a magnetic stirrer. 2.5 mL of 1% $HAuCl_4$ and 7.75 mL of 1% sodium citrate were added. The solution was mixed for 15 minutes. The color of the solution changed from colorless to red. Then, 250 mg of PVP was added to the colloid under mixing. The PVP-coated gold nanospheres were then centrifuged at 20,000 g for 60 min. The supernatant was decanted, and the pellet was resuspended in ethanol. The nanoparticle concentration was about $6 \times 10^{12}\,mL^{-1}$, corresponding to an optical density of 4 as measured in 1-cm cuvette. In the second stage, 10 g of PVP was dissolved in 200 mL of DMSO. After that, 400 μL of gold seeds and 400 μL of 8% $HAuCl_4$ were added to the solution. The reaction of AuNST growth was allowed to proceed overnight. The AuNSTs were cleaned via centrifugation (5000 rpm, 20 min) and resuspended in 200 mL of ethanol.

To tune the LPR wavelength, 50 μL of 8% $HAuCl_4$ was added to 100 mL of initial nanostars colloid in ethanol. The evolution of the AuNSTs was monitored by UV−vis spectroscopy with time and was stopped at the desired wavelength by centrifugation and resuspension in water. The total time of etching was 12 hours. The functionalization of AuNSTs was similar to that of nanorods. A total of 8 samples of AuNST particles were obtained.

**Extinction and TEM measurements**. Extinction spectra were recorded using Specord 250 and Specord S300 spectrophotometers (Analytik Jena, Germany). The measurements were carried out in the 320–1100 nm wavelength range using 2-mm and 10-mm quartz cuvettes. Transmission electron microscopy (TEM) images were obtained using a Libra-120 transmission electron microscope (Carl Zeiss, Germany) at the Simbioz Center for the Collective Use of Research Equipment in the Field of Physico-Chemical Biology and Nanobiotechnology (IBPPM RAS, Saratov). For this purpose, the synthesized gold nanoparticles were deposited on copper grids.

**SERS spectra measurements**. SERS spectra of NBT or Cy 7.5 functionalized AuNR, AuNTs, and AuNSTs were measured with colloids in quartz cuvettes with an optical path length of 1 cm using a Peak Seeker Pro 785 Raman spectrometer (Ocean Optics). The laser power and data collection interval were 30 mW and 10 s, respectively. The laser beam was focused near the cell wall to decrease the internal filter effect.[34] For SERS measurements with a 633-nm laser excitation (in particular, for comparison with reported SERS measurements of AuNST@Cy 7.5 particle[7]), we also used a LabRam HR Evolution spectrometer (HORIBA, Jobin Yvon) with a He-Ne laser at 632.8 nm.

**Electromagnetic simulations**. The extinction spectra of AuNRs were simulated by using T-matrix codes developed previously.[35] The bulk dielectric function of Au was taken from Olmom et al.[36] All details concerning the nanometer-size correction can be found in Ref.[37] Specifically, the surface scattering constant of Au conductive electrons, $A_s = 0.33$, was taken according to the previously reported recipe.[38] The refractive index of water was calculated as described in Ref.[38]. COMSOL simulations of extinction spectra for AuNTs are described in detail in the Supporting Information file, Sections S2.3-S2.5.

The orientation and surface averaged SERS enhancement factor as a function of AuNR aspect ratio and LPR wavelength was calculated as described in detail in Section S 2.7. Briefly, COMSOL calculations were carried out by the finite element method (Comsol 5.3, Wave optics module) within the 3D full-wave approach in terms of the scattered field formulation. The boundaries of a spherical computational domain were limited by PML. The field continuity conditions were satisfied on all surfaces. The surface-averaged SERS enhancement factor was calculated by integrating the product of local field intensities at the laser and Raman frequencies, $|EF(\omega_L)|^2 |EF(\omega_R)|^2$ [8] (Eq. S12). Due to the axial symmetry of AuNRs, the orientation averaging over the particle orientations was replaced with the averaging over the polar angle between the incident field and the long particle axis. For simulation details and COMSOL codes, the readers are referred to Section S2.7.

**RESULTS AND DISCUSSION**

**Gold nanorod synthesis and characterization**. Figure 1 summarizes TEM and spectrophotometry data for all 13 AuNR samples. The average length, diameter, and aspect ratio of as-prepared AuNRs from TEM images were $L = 76.6 \pm 8.1 \, \text{nm}, d = 13.4 \pm 0.89 \, \text{nm}$, and $AR = 5.08 \pm 0.60$, respectively (Table 1). The initial longitudinal LPR was located at 1017 nm.

Slight broadening and high ratio of the major to minor peak amplitudes indicate the high quality of the sample with a low percentage of byproduct particles (less than 0.5%; see Figure S1A).[39] Typically, there were no more than 4 nanospheres per 800 nanorods in TEM images. This high quality is explained by the separation of nanorods obtained with the hydroquinone protocol.[25]

The etching of AuNRs was performed by the addition of HAuCl₄ to the solution of nanorods in 0.1 M CTAB. According to the mechanistic study by Liz-Marzán group,[23] the dissolution of AuNRs is governed by reaction

$$AuCl_4^- + 2Au^0 + 2Cl^- \rightleftharpoons 3AuCl_2^-. \tag{1}$$

Without CTAB, the equilibrium constant of reaction (1) is only $1.9 \times 10^{-8}$; thus, the reduction of $Au^{3+}$ to $Au^0$ is negligible. However, adding 0.1 M CTAB and formation of $AuCl_2^- + CTAB$ complexes shifts the equilibrium constant to 93.8 and enables the complete consummation of two $Au^0$ moles for each $Au^{3+}$ mole added. Because of the dense adsorption of CTAB on the side nanorod surface, the above mechanism ensures the directional dissolution of the nanorod ends. As a result, the nanorod morphology and diameter remain unchanged while the length, aspect ratio, and LPR wavelength gradually decrease. This procedure is very flexible because of the linear dependence between the molar concentrations of $[Au^0]$ and $[Au^{3+}]$ (see Fig. 1 in Ref.[23]). Thus, by variation of the added $[Au^{3+}]$ amount, it is possible to tune the LPR wavelength to any desired wavelength within 1100-600 nm with accuracy better than 5 nm.[24]

During the etching process, the thickness of AuNRs in all samples is close to the average value of 13.8 nm with a weak increasing tendency. In comparison, the average length decreases from 76 to 22 nm, thus decreasing the aspect ratio from 5.4 to 1.6 (Figure 1 E). Accordingly, the major LPR peak wavelength gradually decreases from 1017 nm to 580 nm (Figure 1F). Note that for most shortened AuNRs, the extinction peak splits into two modes in agreement with the T-matrix simulations (Figure S5).

Additional TEM images and statistical distributions are in the Supporting Information (SI) file (Section S1, Figures S1-S4). Table 1 summarizes the average geometrical parameters and longitudinal LPRs of AuNRs.

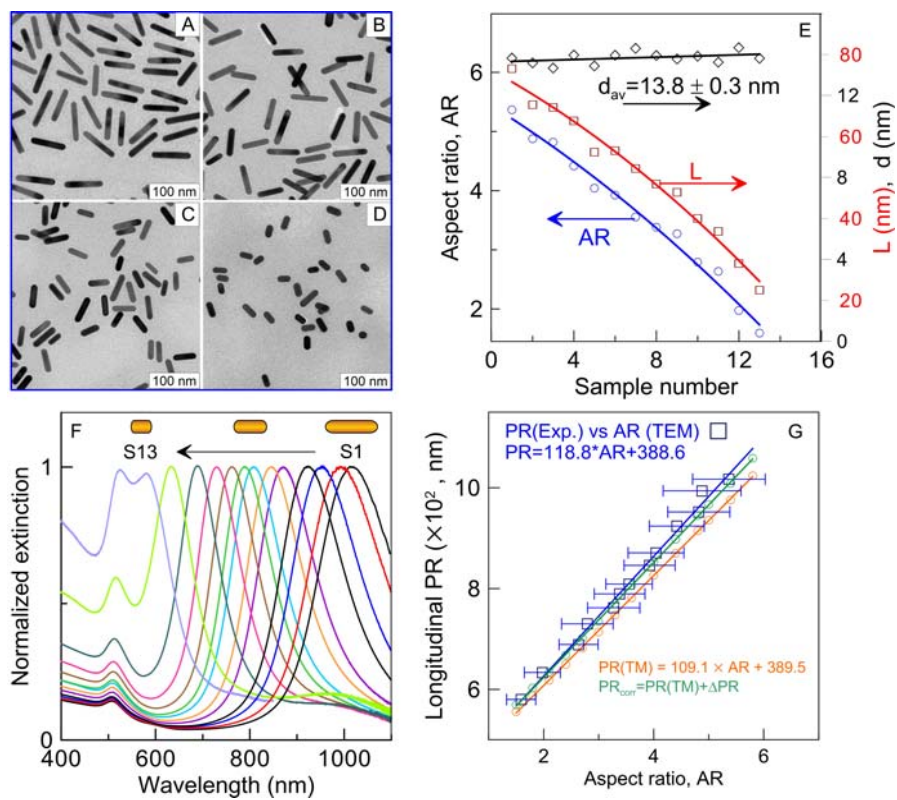

**Figure 1**. TEM images of samples AuNR 1 (A), 4 (B), 8 (C), and 12 (D). (E) – plots of the AuNR length, diameter, and aspect ratio as a function of the sample number. (F) – extinction evolution spectra of etched samples 1 – 13. (G) – experimental LPR wavelength as a function of the TEM measured aspect ratio (blue), T-matrix calculated plot LPR(AR) for bare AuNRs (orange), and that corrected for CTAB layer plasmonic shift (green).

**Table 1**. Aspect ratio, length, diameter, and longitudinal LPR wavelengths of etched nanorods.

| Sample | AR | L, nm | d, nm | LPR, nm |
|---|---|---|---|---|
| 1(initial) | 5.37±0.66 | 76.6±8.1 | 13.8±0.76 | 1017 |
| 2 | 4.88±0.71 | 67.8±10.3 | 13.6±1.1 | 994 |
| 3 | 4.82±0.56 | 67.1±7.7 | 13.3±0.79 | 952 |
| 4 | 4.49±0.45 | 63.9±7.5 | 14.0±0.93 | 924 |
| 5 | 4.04±0.51 | 56.3±7.8 | 13.5±0.83 | 871 |
| 6 | 3.93±0.46 | 56.6±6.8 | 14.0±0.78 | 846 |
| 7 | 3.55±0.42 | 52.2±6.6 | 14.3±0.95 | 809 |
| 8 | 3.38±0.46 | 48.5±6.7 | 14.0±0.84 | 789.5 |
| 9 | 3.27±0.48 | 46.4±7.0 | 13.8±0.75 | 762 |
| 10 | 2.79±0.47 | 40.1±7.4 | 13.9±0.90 | 730 |
| 11 | 2.64±0.35 | 36.9±5.3 | 13.6±0.68 | 689 |
| 12 | 1.98±0.33 | 29.0±5.2 | 14.3±0.85 | 633.5 |
| 13 | 1.59±0.26 | 22.5±3.8 | 13.8±0.65 | 580 |

Figure 1G shows a well-known linear correlation between the AuNR aspect ratio and the longitudinal LPR wavelength.[40] However, T-matrix simulations with TEM-derived geometrical parameters of AuNRs (red line in Figure 1G) give the linear fit

$$PR(\text{TM}) \equiv \lambda_{\max}(\text{TM}) = 109.1 \times AR + 389.5, \quad (2)$$

that does not agree with experimental dependence (blue fit line). Similar disagreement was also observed for the simplest electrostatic simulations.[41] This disagreement can be related, at least in part, to the LPR redshift caused by the CTAB coating of the bare AuNRs. Indeed, COMSOL simulations (SI, Figure S6) demonstrate notable red-shifting of extinction spectra.

Two CTAB shell parameters determine the LPR shift: the shell thickness, $t_s$, and its refractive index. According to the Powerful Online Chemical Database,[42] the refractive index is $n_s$ =1.526. For CTAB solutions, Ekwall et al.[43] reported the refractive index increment of 0.150 mL/g, which, together with the specific volume of 0.984 g/mL and the refractive index of water 1.334, gives $n_s$ =1.486. Two other papers[44][45] indicated a somewhat smaller value of 1.435. In our simulations, we used the average value $n_s$ =1.48.

The CTAB shell thickness has been measured in AuNR colloids with SAXS[46,47] and SANS[48] techniques to give the average value of 3.2 nm.[49] However, the single-particle measurements with the electron energy loss spectroscopy in an aberration-corrected scanning transmission electron microscope (STEM-EELS) revealed a somewhat smaller thickness of 2.4 nm. Here, we used an average value of 3 nm to characterize the shell thickness of CTAB molecules adsorbed on the side surface and tips of AuNRs. We assumed a homogeneous distribution of CTAB molecules on the nanorod surface instead of an asymmetric dipole-like distribution discussed in Ref.[50]

From COMSOL simulations, we derived the following correlation between the CTAB-coating plasmonic shift and the aspect ratio of nanorods

$$\Delta \lambda_{PR} \equiv \lambda_{max}(s=3\text{nm}) - \lambda_{max}(s=0) = 4.1 \times AR + 7.12 . \quad (3)$$

After inserting this correction into Eq. (2), we get the corrected equation (green line in Figure 1G)

$$\lambda_{max}(\text{TM}) = 113.21 \times AR + 396.6 , \quad (4)$$

which is in excellent agreement with the experimental linear fit (blue line). Thus, including the CTAB shell in the simulation model allows for quite satisfactory prediction of the LPR peak position as a function of the aspect ratio.

**SERS measurements for AuNRs.** Before discussing the measurement results, it is instructive to look at the results of modeling the dependence of the SERS enhancement factor (EF) on the plasmon resonance of nanorods (or, equivalently, on their aspect ratio) for AuNR@NBT (line 1343 cm$^{-1}$) and AuNR@Cy7.5 (line 940 cm$^{-1}$) particles for two excitation laser wavelengths of 633 and 785 nm (Figure 2). Several important notes can be drawn from these simulations.

(1) The orientation averaging decreases the maximal EF roughly five times (compare plots in panels A and B). This observation agrees with the simplest forth-power law because the average value of the local field should be proportional to $\langle \cos^4 \theta \rangle = 1/5$, where $\theta$ is the polar angle between the incident electric field and the particle axis;

(2) The calculated SERS profiles demonstrate two split maxima corresponding to the laser and Raman frequencies in the product $|E(\omega_L)|^2 |E(\omega_R)|^2$ (see Eq. S12);

(3) In agreement with the $\lambda_{LPR} = (\lambda_{ex} + \lambda_{RS})/2$ [9, 17] rule, the laser wavelength is blue-shifted with respect to the optimal LPR wavelength;

(4) EFs for 785-nm excitation are expected to be 5-7 times higher compared to those for 633-nm laser;

(5) The electromagnetic contribution to the enhancement of SERS line 940 cm$^{-1}$ of AuNR@Cy7.5 particles is higher than that for line 1343 cm$^{-1}$ of AuNR@NBT particles. Together with higher Raman cross sections of Cy7.5 molecules, one could expect a much more intensive SERS response from AuNR@Cy7.5 particles than that from AuNR@NBT ones;

(6) Owing to the Raman shifts between $\omega_L$ and $\omega_R$ frequencies, the fourth-power law is violated within 600-700 and 800-900 nm spectral bands where EF is roughly constant for 633-nm and 785-nm lasers, respectively. This observation has been confirmed previously for thick 26-nm Au nanorods experimentally (785 nm) and numerically.[8]

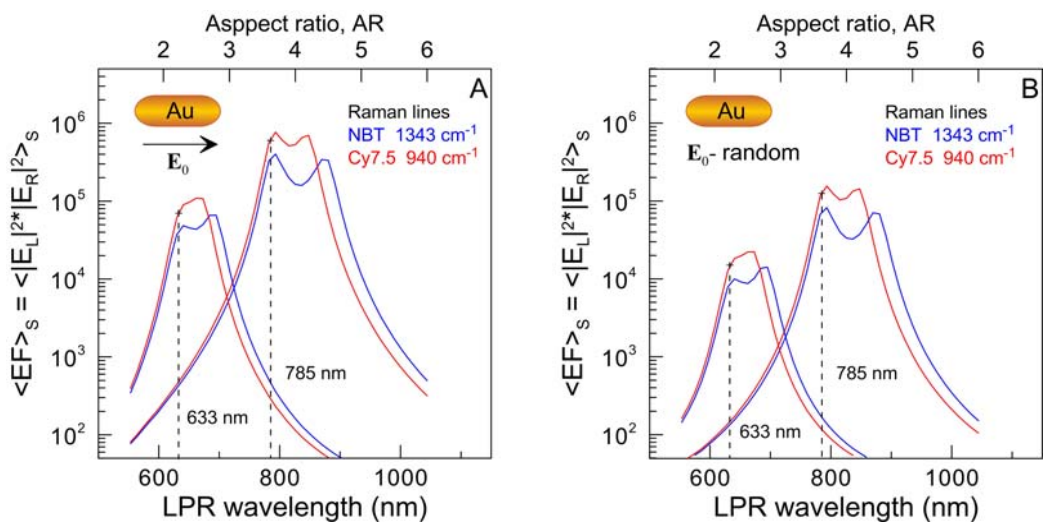

**Figure 2**. Dependence of the surface averaged SERS enhancement factor for the intensities of Raman lines 940 cm$^{-1}$ (red curves, AuNR@Cy7.5 particles) and 1343 cm$^{-1}$ (blue curves, AuNR@NBT particles) on the LPR wavelength at longitudinal (A) and random (B) orientation of the exciting electric field. The plots in panel B show EFs averaged over the particle surface and their random orientations. Note a five-fold decrease in EF maxima due to random orientations. The vertical dashed lines show the laser wavelengths of 633 and 785 nm, and the upper abscissa shows the aspect ratios of the nanorods.

Now, we proceed to experimental data. Figure 2A shows the LPR-wavelength dependence of the SERS peak at 940 cm$^{-1}$ of AuNR@Cy7.5 samples measured at 785 nm laser excitation. Additional SERS spectra are shown in Figures S7 (AuNR@NBT, 785-nm laser), S8 (AuNR@NBT, 633-nm laser), and S9 (AuNR@Cy7.5, 785-nm laser). The maximal SERS peak is observed for a sample with an LPR wavelength of 762 nm, which is slightly blue-shifted with respect to the laser wavelength of 785 nm. As we discussed in the Introduction section, the electromagnetic theory[9] and some experimental observations[11, 16, 17] demonstrated the maximal SERS responses for samples with LPR wavelength slightly redshifted compared to the laser wavelength. At first glance, our data in Figure 3A appear to be at odds with the rule $\lambda_{LPR} = (\lambda_{ex} + \lambda_{RS})/2$.[9, 17] However, Sivapalan et al.[51] showed that the resonance absorption of light at LPR wavelength results in notable extinction of both excitation and Raman scattered light traveling in AuNR suspension. This extinction leads to observing maximal SERS response for samples with a smaller aspect ratio (and LPR wavelength) than expected from a thin film substrate or single particle measurements. Thus, we conclude that the resonance extinction in colloids can explain the somewhat unexpected blue-shifted optimal LPR in Figure 1A. In other words, this is a manifestation of the inner filter effect.[34]

A similar blue-shifted optimal LPR for maximal SERS is demonstrated in Figure 3E for NBT functionalized AuNRs. Again, the maximal SERS response is observed for sample 5 with an LPR of 762 nm. We previously observed a small blue-shifting of optimal LPR for NBT-functionalized AuNRs with a thicker diameter of 26 nm.[8] However, as distinct from those measurements and data in Figure 3A, the NBT-functionalized AuNRs with a diameter of 13.8 nm in Figure 3D demonstrate a much weaker SERS response with peak intensity 30 times smaller compared to that for AuNR@Cy7.5 sample 5 in Figure 3A and our previous data for 26-nm AuNR@NBT nanorods.[8] Two reasons can explain this discrepancy. First, the Raman cross-section of Cy7.5 molecules is greater than that of NBT[52], thus explaining the difference between the data in Figures 3A and 3E. On the other hand, the difference between the maximal SERS peaks observed for thick[8] and thin (Figure 3E) nanorods can be related, at least in part, to the different surface covering of AuNR with Raman reporters.

In panel F, we summarize data for normalized SERS intensities measured for Cy7.5 and NBT functionalized AuNRs with a diameter of 13.8 nm (red circles and blue stars) as well as our previous measurements[8] for thick 26-nm AuNRs functionalized with NBT (green triangles). Two important notes are in order here. First, all experimental maxima are slightly blue-shifted compared to the laser excitation wavelength, following the above explanation.

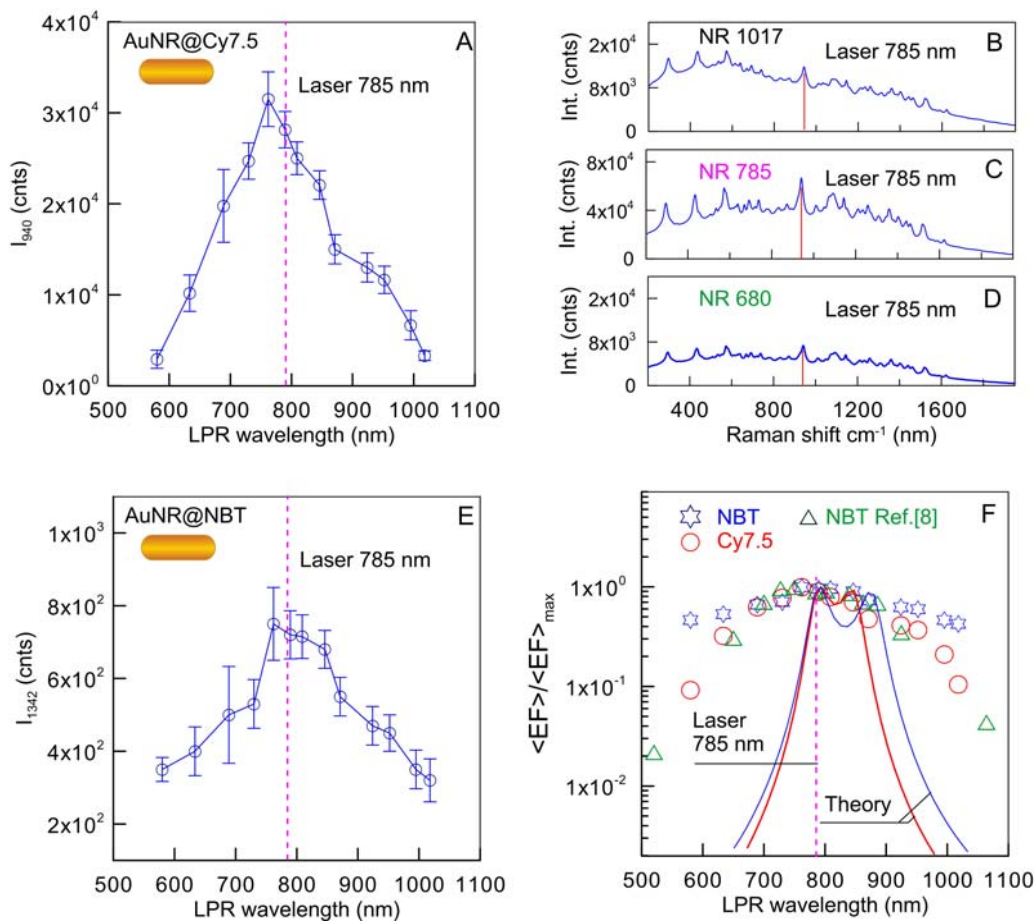

**Figure 3**. Dependence of SERS peak intensities $I_{940}$ (A, AuNR@Cy7.5) and $I_{1343}$ (E, AuNR@NBT) on the LPR wavelength as measured at 785-nm laser excitation for 13 AuNR samples. Note a slight blue shift of maximal peak intensities for the laser wavelength 785 nm. Panels B-D illustrate untreated SERS spectra of AuNR@Cy7.5 particles for 3 samples with AuNR LPRs of 1017, 785, and 680 nm (more spectra are shown in Figure S9). Panel F summarizes data for normalized SERS intensities measured for Cy7.5 and NBT functionalized AuNRs with a diameter of 13.8 nm (red circles and blue stars). Green triangles reproduce measurements for thick 26-nm AuNRs functionalized with NBT (data of Ref.[8]). The solid curves

show the theoretical normalized enhancement factor for Cy7.5 (red) and NBT (blue) Raman lines. These plots were calculated for 785-nm laser and 13.8-nm AuNRs with surface and orientation averaging as described in Section 2.7.[8]

Second, the experimental SERS profiles are much broader than those predicted by electromagnetic simulations. Regarding these simulations, we note ~50 nm red-shifting of the theoretical SERS profile with respect to the laser wavelength, in agreement with the simplest $\lambda_{LPR} = (\lambda_{ex} + \lambda_{RS})/2$ rule. Further, there is a spectral range within 750-900 nm in which the theoretical SERS profile varies weakly in agreement with experimental data. Nevertheless, we emphasize strong disagreement between the simulated and measured broadening of the SERS profiles. This disagreement can be attributed to various origins, including many simplifications adopted in simulations[8] and known problems in calculations of the fundamental enhancement factor from experimental SERS spectra.[2]

To evaluate the excitation wavelength's effect, we measured the SERS profile for the same 13 AuNR@NBT samples using the LabRam HR Evolution spectrometer operating at a laser wavelength of 633 nm. The measured SERS spectra are shown in Figure S8, and Figure S10 shows the dependence of the peak intensity at 1343 cm$^{-1}$ on the LPR wavelength of samples. Since only one LPR wavelength of sample 13 (580 nm) was shorter than the laser wavelength of 633 nm, almost all samples showed similar small peak intensities, except for three samples with minimal wavelengths. We consider these data as rough estimates since, in this case, the measurements near the laser wavelength could be strongly distorted by the extinction effects of the exciting and scattered light.[51] These distortions can explain the maximum signal for samples with a resonance wavelength smaller than the laser wavelength. For more accurate estimates, additional studies of the issue are needed.

**Characterization and simulations of gold nanotriangles**. Figure 4A shows TEM images of the initial and etched AuNT particles. The shape of initial particles can be approximated by triangles with rounded tips (see Figure S11 for additional TEM images). The initial sample 0 contained a notable amount of byproduct particles (Figures S11 A and B) and was separated additionally. Further, the extinction spectrum of the as-prepared initial sample had very intense plasmonic peak extinction (~2.8), thus exhibiting notable fluctuations (gray spectrum in Figure 4F).

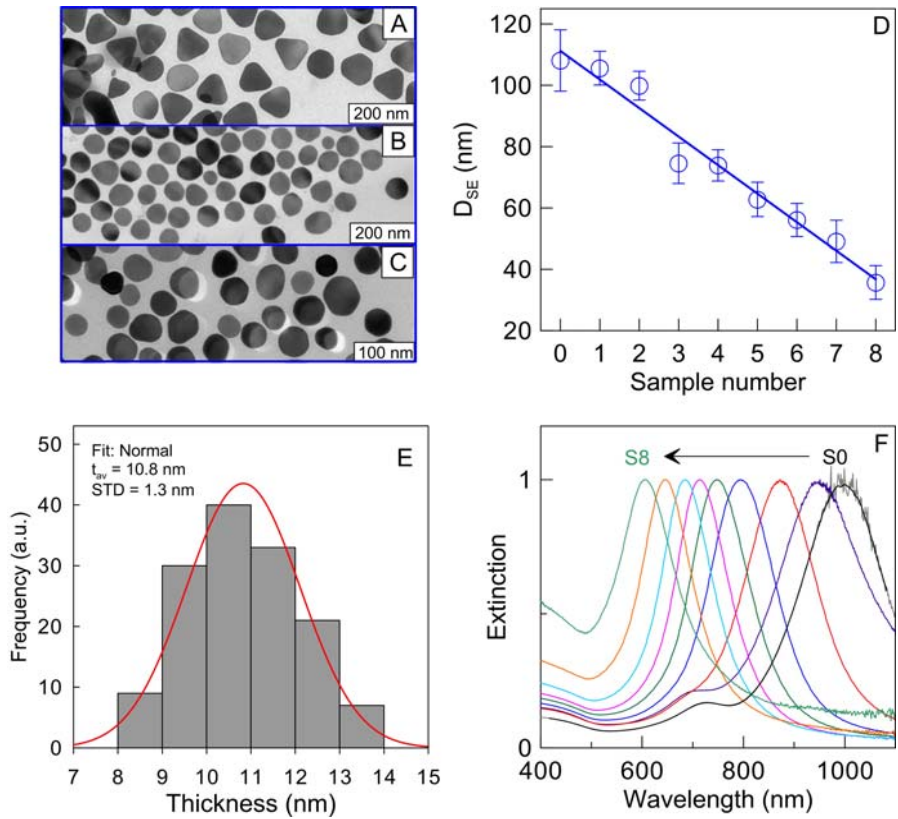

**Figure 4**. TEM images of initial AuNTs (A, LPR at 950 nm), sample 4 (B, LPR at 749 nm), and sample 8 (C, LPR at 607 nm). Panel D illustrates the decrease in the average particle size (regarding the surface equivalent diameter $D_{SE} = 4S/\pi$). Panel E shows the statistical histogram and the normal Gaussian fit (red) for the TEM-derived thickness with the average $t_{av} \simeq 11$ nm

value. The right bottom panel F shows the evolution of extinction spectra during the etching process for samples 0-8 (from the right to the left). A gray line shows the spectrum of the initial sample 0, and the black spectrum shows the running average approximation to smooth the fluctuations.

According to TEM data (Figure S12), the AuNT thickness of samples 1-8 is approximately constant. It equals 11 nm as determined from the thickness of random particle stacks and divided by the number of particles in the stack.

**Table 2.** The average surface-equivalent diameter ($D_{SE}$) and LPR wavelengths of etched nanotriangles.

| Sample | 0 | 1 | 2 | 3 | 4 | 5 | 6 | 7 | 8 |
|---|---|---|---|---|---|---|---|---|---|
| $D_{SE}$ (nm) | 108.1±10.0 | 105.6±5.5 | 99.8±6.6 | 74.6±6.6 | 73.9±5.1 | 62.8±5.6 | 56.1±5.4 | 49.1±6.9 | 35.7±5.5 |
| LPR (nm) | 1010 | 950 | 873 | 795 | 749 | 714 | 685 | 646 | 607 |

As the nanoparticles are etched, their mean surface diameter decreases from 108.1 nm to 35.7 nm (Figure 4D and Table 2), and the plasmon resonance shifts from 1010 nm to 607 nm (Figure 4F). As a result of etching, the shape of the particles gradually changes from triangular to the shape of irregular disks (Figure 4B). Note the presence of a multipole shoulder at about 700 nm in the extinction spectrum of the initial sample with triangular particles.

Figure 5A shows a model for describing the size and shape of a particle in the etched sample 1 using the length of the described triangle side ($L$) and the rounding radius of its tip ($R$). More detailed information concerning the characterization of AuNT size and shape is given in Section S2.2 (SI file).

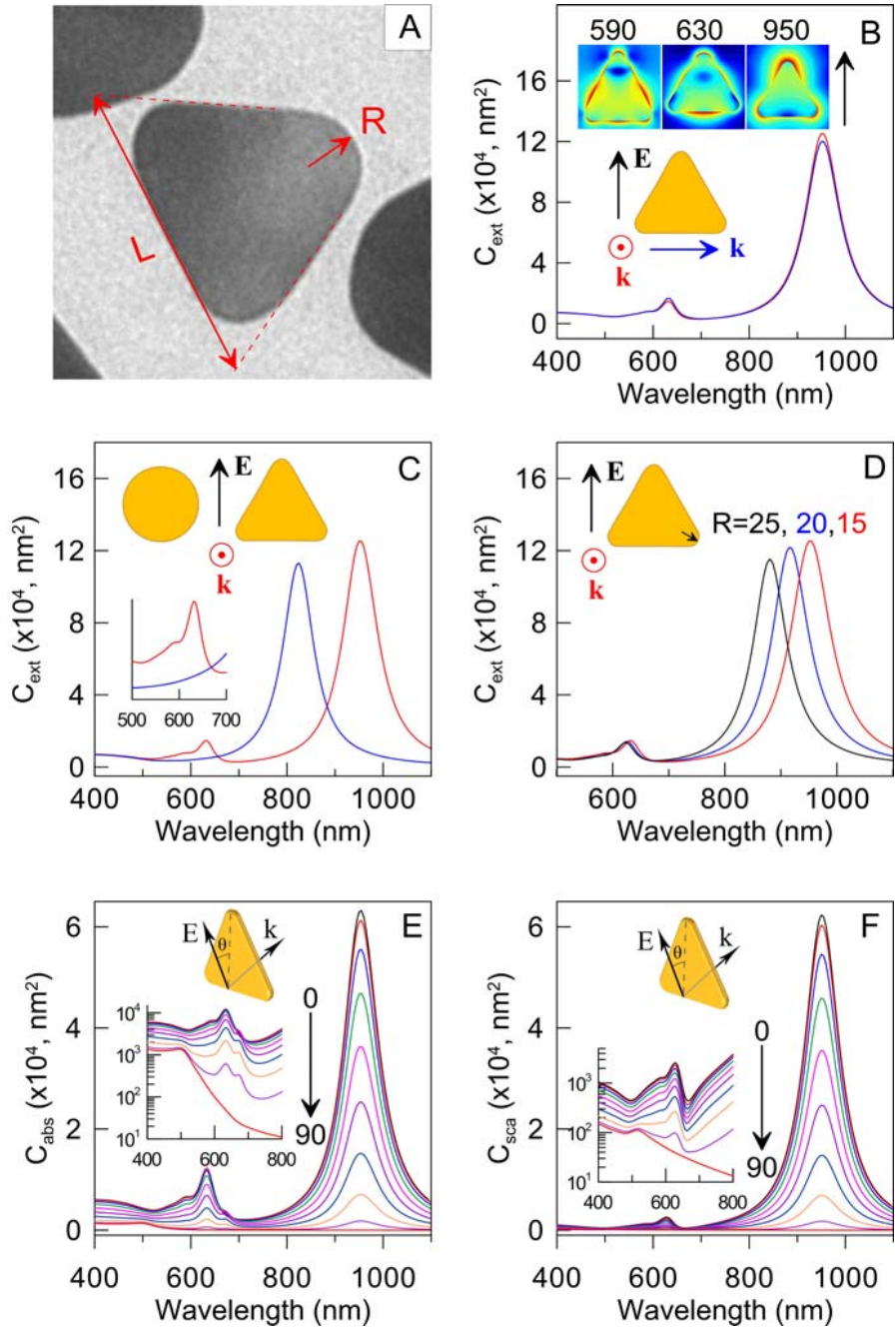

**Figure 5.** (A) TEM image of a gold nanotriangle (sample 1) and a model for describing its size and shape in terms of the length *L* of the side of the circumscribed triangle and the vertex radius *R* (see also Section S2.3). (B) Extinction spectra are calculated for the electric field in the particle plane and the wave vector in-plane and out-of-plane incidence. The inset shows the local field

distributions near the particle for three wavelengths corresponding to the excitation of resonances in the extinction spectrum. Calculation using the COMSOL package for $L$=160 nm and $R$=15 nm values obtained from TEM analysis. (C) The short-wavelength shift of the spectrum of the triangle upon transformation into a disk of equal area. (D) The long-wavelength shift of the extinction spectrum with decreasing tip rounding radius. (E) Evolution of the absorption spectrum as the angle between the field vector and the nanoparticle plane increases from 0 to 90 degrees. (F) Same as in panel E for the scattering spectrum. The insets in panels E and F show multipole resonances in the 500–700 nm region.

Calculations by the Finite Element Method (COMSOL package) showed that the extinction spectrum upon excitation in the particle plane is practically independent of the orientation of the wave vector (blue and red spectra in Figure 5B and Figure S14). This important observation leads to efficient approximations for calculating the optical cross sections averaged over random orientations relevant to AuNT colloids. In particular, the average extinction cross section can be calculated by the following approximation (Section S2.4)

$$\langle C_{ext} \rangle = \frac{C_{ext}^z(\mathbf{k}_y) + C_{ext}^z(\mathbf{k}_x) + C_{ext}^y(\mathbf{k}_z)}{3}, \qquad (5)$$

where the nanotriangle is in the (x,z) plane with one side directed along the x-axis (Figure S12). In Eq. (5), the upper superscript stands for the incident electric field orientation along the z and y axes, and the wave vectors are directed along the y, x, and z axes, respectively. The accuracy of approximation (5) is estimated in Section 2.4. In electrostatic approximation, the orientation averaged cross section equals

$$\langle C_{ext} \rangle = \frac{C_{ext}^x + C_{ext}^y + C_{ext}^z}{3} \simeq \frac{2C_{ext}^z + C_{ext}^y}{3}, \qquad (6)$$

where the numerically proved approximation $C_{ext}^x \simeq C_{ext}^z$ has been used. Although Eq. (6) has been used as the right numerical recipe for nanotriangles[53] and nanoshells,[54] we emphasize here that the relation (6) is, in fact, only an electrostatic approximation even though the cross section themselves are calculated accurately with any numerical method (e.g., by FDTD).

The major LPR peak is located at 950 nm for a particle model with $L = 160$ nm and $R = 15$ nm, in close agreement with the experimental spectrum for the etched sample 1 (Figure S14). The upper inset in Figure 5B illustrates the near-field distribution at three resonance wavelengths. In addition to the primary dipole resonance at a wavelength of 950 nm with the near-field enhancement near the collinear vertex of the triangle (upper inset on the right), there is also a resonance at a wavelength of 630 nm with the field on the lower side of the triangle (center of the upper inset) and a weak resonance with a field concentration near the sides (upper inset on the left). When the triangle is transformed into a disk, its spectrum shifts to short wavelengths, and the multipole resonances degenerate due to particle symmetry. In particular, Figure 5C illustrates the disappearance of two multipole peaks at 590 and 630 nm when a triangle is transformed into a disk.

The next important parameter is the vertex radius $R$. Specifically, with an increase in the vertex radius, the LPR peak shifts to the short wavelengths, while the tip sharpening results in the redshifted LPR (Figure 5D). Finally, the LPR amplitude strongly depends on the orientation of the exciting field to the nanoparticle plane. With a gradual change in the angle of inclination of the exciting incident field to the particle plane, strong dipole resonances of absorption (E) and scattering (F) decrease and eventually disappear upon perpendicular excitation. Note that the absorption spectrum in the 500–700 nm spectral band has a more complex multipole structure than the scattering spectrum (see insets in Figures 5 E and F).

In Section S2.4, we calculated the extinction spectra of AuNT ensembles using a three-fraction model that includes contribution from AuNT triangles, disks, and elliptic disks. Table S3 provides information about geometrical models of particles and their numerical concentrations (in terms of percentages). This information describes a gradual transformation of AuNTs into disks and smaller elliptical disks during the etching process. Specifically, the dominant nanotriangle fraction (~80%) is transformed into the dominant disk fraction (72 %) and a minor elliptic disc fraction (28 %). COMSOL-simulated extinction spectra are shown in Figure S17 and closely resemble the time evolution of experimental spectra (Figure 4F), except for notable narrow spectra peaks and their precise positions. These deviations from experimental spectra can be attributed to a simplified model of complex real AuNT ensembles.

**SERS measurements for AuNTs.** Measurement of SERS spectra for AuNT@NBT and AuNT@Cy7.5 samples (Figures S18 and S19) showed that the peak intensities of the most intense lines (940 cm$^{-1}$ for Cy7.5 and 1343 cm$^{-1}$ for NBT) are maximal when the plasmon resonance of the nanotriangles is close to the laser wavelength of 785 nm (Figure 6A). Specifically, the SERS peak of AuNT@NBT particles is slightly redshifted with respect to the laser wavelength, and its intensity is almost 40 times lower than that for Cy7.5. These observations agree with those for AuNR@NBT data shown in Figure 3E. However, unlike Figure 3A, the SERS profile for AuNT@Cy7.5 in Figure 6A is slightly redshifted in agreement with classical EM theory. This difference is probably related to the different geometry of the particles and a weaker inner filter effect.

Figure 6B shows the plasmonic SERS profile for AuNT@NBT particles at 633 nm laser excitation. The optimal LPR wavelength that maximizes SERS response is slightly redshifted compared to the laser wavelength, as predicted by theory. Further, the SERS intensities are almost one order lower than those for AuNT@Cy7.5 particles (red line in panel A). Similarly to the data for functionalized AuNRs, we note that both SERS profiles in Figure 6 are much broader than those predicted from EM simulations.

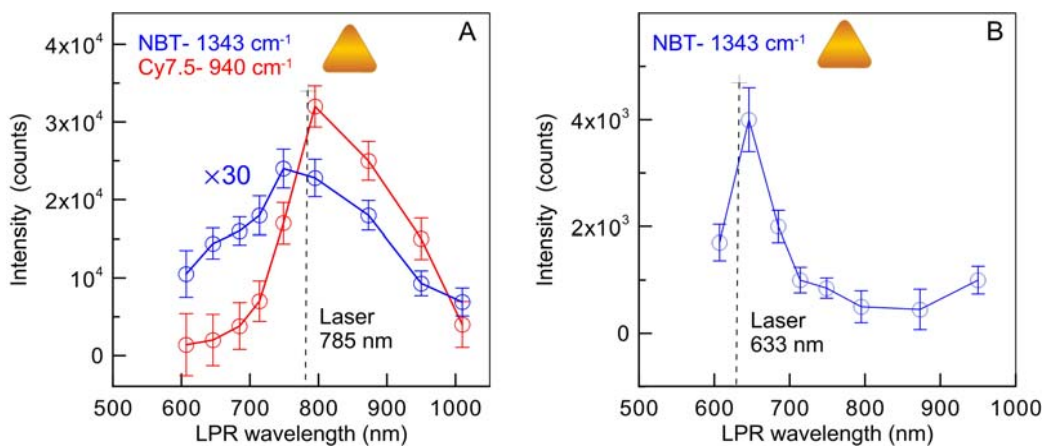

**Figure 6**. Dependence of SERS intensities $I_{1341}$ (NBT) and $I_{940}$ (Cy7.5) on the LPR wavelength of AuNTs for laser excitation at 785 nm (A) and 633 nm (B). Note that experimental points for AuNT@NBT intensities in panel A are multiplied by 30.

**Characterization and SERS measurements for AuNSTs**. AuNST samples were prepared by a method described previously[7] with some modifications. The initial non-etched AuNSTs have an LPR wavelength of 838 nm, and after etching, we fabricated 8 samples with LPR wavelengths listed in Table S4 (see also histograms in Figs. S21-S23). The LPR wavelength is determined mainly by the tip sharpness, whereas the core size gives a minor contribution.[7, 37, 55] The tip sharpness is determined by three correlated parameters, namely the tip length, angle, and end diameter $D_{tip}$ (Figure 7E). According to an experimental study,[7] the tip angle can be considered the main combined parameter for LPR tuning, whereas other parameters give only an additional minor contribution. With a decrease in the tip-end diameter $D_{tip}$ values, the LPR wavelength shifts to red, as was demonstrated for Au nanoantennas by Tsolulos and Fabris.[55] For our nanostars, this effect is less significant. During etching, the AuNST tips lose their sharp morphology, shorten, and smooth out (Figures 7A, B). Finally, the tips disappear entirely, and the etched nanostars look like nanospheres with a few smooth protrusions (Figure 7C). These morphological changes are accomplished by a continuous shift of LPR wavelength from 838 nm to 565 nm (Figures 7D and F). The plots in panel F illustrate the morphological mechanisms behind the LPR tuning of etched AuNSTs. In agreement with previous conclusions,[7] the main effect comes from the tip angle and length, whereas the core diameter gives only a minor contribution.

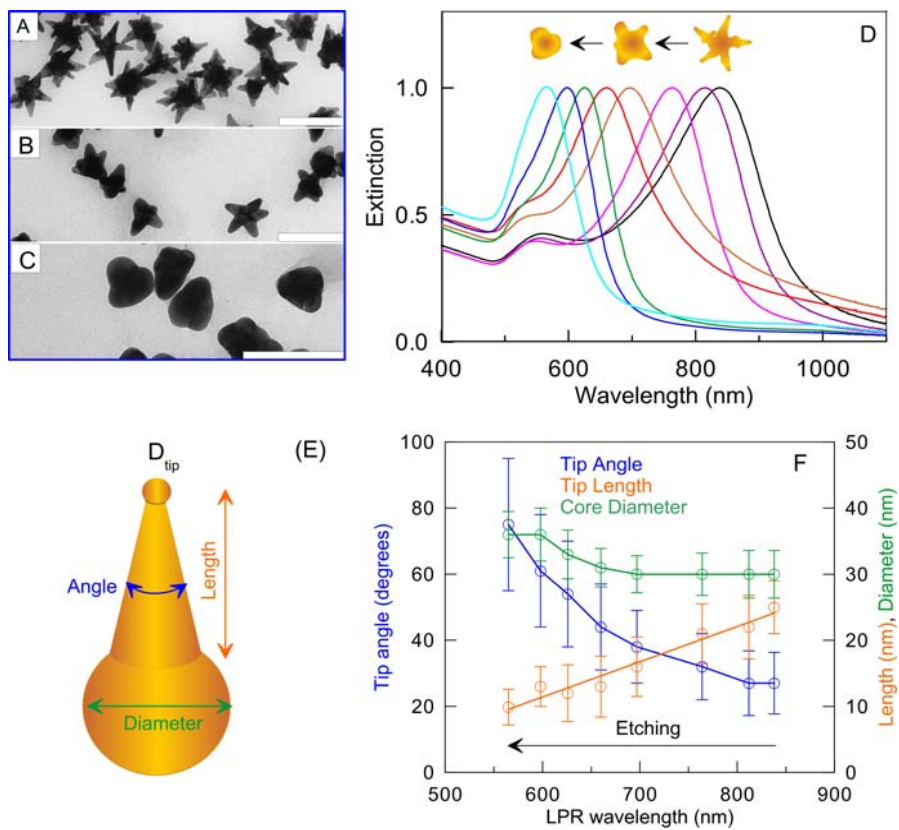

**Figure 7**. TEM images of AuNST with LPR wavelengths 838 nm (A, initial sample), 660 nm (B), and 565 nm (C) and evolution of extinction spectra during chemical etching. The scale bars are 100 nm. Panels E and F show morphological parameters of etched AuNSTs (tip angle,

length, and core diameter) and their correlation with the LPR wavelength that decreases during the chemical etching from 838 to 565 nm.

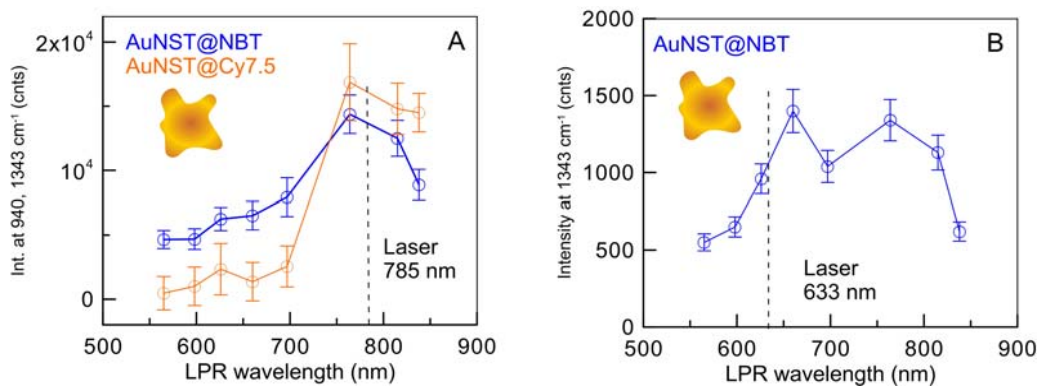

**Figure 8**. The dependence of SERS intensities on the LPR wavelength of AuNSTs measured for two Raman lines 940 cm$^{-1}$ (Cy7.5, orange) and 1343 cm$^{-1}$ (NBT, blue) at 785-nm (A) and 633-nm (B) laser excitation.

Figure 8 shows the SERS intensities as a function of the LPR wavelength measured for AuNCT@Cy7.5 and AuNST@NBT particles with Raman lines 940 cm$^{-1}$ and 1343 cm$^{-1}$, respectively. For the 785-nm laser, the maximal SERS response is observed at LPR wavelength, slightly blue-shifted with respect to the laser wavelength. Similarly to previous measurements with AuNRs and AuNTs, we attribute this blue shit to the inner filter effect. For the 633 nm laser, the maximal SERS intensity was recorded for the LPR wavelength of 660 nm, in agreement with the electromagnetic prediction for redshifted optimal LPR.[1] By contrast to the strong 35-fold difference between maximal SERS intensities for AuNR(NT)@Cy7.5 and AuNR(NT)@NBT particles (Figures 3 and 6), here we observe more or less similar intensity for both Raman molecules adsorbed on AuNSTs. As the Raman cross section of Cy7.5 molecules is much greater than that for NBT ones, we can attribute the similarity of plots in Figure 7E to the small surface density of Cy7.5 molecules adsorbed on AuNSTs.

Finally, we have to note strong broadening SERS plasmonic profiles compared to what one would expect from theoretical predictions. Moreover, at 633-nm laser, we recorded almost constant SERS response from AuNST@NBT particles within the 640-840 LPR range. This observation is in line with previous measurements for thick 26-nm Au nanorods functionalized with NBT molecules.[8]

We conclude this section by pointing to an interesting numerical study of the planar nanostar structures, which provide the maximum SERS response for a given excitation wavelength.[56]

**Conclusions.**

In this work, we synthesized, characterized by spectroscopy and TEM methods, and studied the dependence of the SERS electromagnetic amplification factor on the plasmon resonance wavelength of three types of colloidal nanoparticles (AuNRs, AuNTs, AuNSTs) functionalized by two types of Raman molecules and excited by two laser wavelengths of 785 and 633 nm. Comparing the simulated extinction spectra of AuNRs with the T-matrix and COMSOL methods, we showed that the theoretical dependence of the LPR wavelength on the aspect ratio does not agree with the TEM-derived data if the 3-nm CTAB layer on the nanorod surface is not included in simulations. Using the chemical etching method, we synthesized 13 AuNR samples with LPR wavelengths from 1017 to 580 nm, functionalized them with Cy7.5 and NBT molecules, and measured their SERS spectra upon excitation with 633 nm and 785 nm lasers. For comparison with the measurement data, the plasmon profiles of the SERS enhancement factor EF were calculated by averaging the product of the field intensities of the local field at the laser and Raman wavelengths over the surface of particles and their random orientations. Qualitatively, the simulation results agree with the measurements. In particular, the theoretical maximum EF for AuNR@Cy7.5 particles is two times greater than the maximum for

AuNR@NBT. Together with a higher Raman cross section of Cy7.5 (compared to NBT), this explains the experimental 35-fold difference between SERS EFs for AuNR@Cy7.5 and AuNR@NBT particles. The main difference between measurements and simulations is that the experimental maximum EF upon excitation with a 785 nm laser is observed at a wavelength LPR less than the laser wavelength, while in theory, it is vice versa. Finally, the width of the experimental plasmon gain SERS profile is much broader than the calculated one.

For gold nanotriangles, efficient methods have been developed for averaging the integral optical cross sections over random particle orientations. The results of measurement and analysis of the SERS spectra for AuNT@Cy7.5 and AuNT@NBT particles upon excitation by 633 nm and 785 nm lasers are largely similar to the conclusions obtained for nanorods. In the case of gold nanostars, the main difference from the results for previous experimental models is a small difference between the maximum EF values for AuNT@Cy7.5 and AuNT@NBT particles. The rest of the conclusions apply to these particles as well.

In summary, our experimental and theoretical data provide new insight into the physical mechanisms behind the plasmonic enhancement of Raman signals by nanoparticles with different morphologies.

**Acknowledgments**

This research was supported by the Russian Science Foundation (project no. № 23-24-00062). The work on computer simulations was funded by the Ministry of Science and Higher Education of the Russian Federation as a state assignment for the Saratov Scientific Centre of the Russian Academy of Sciences (research topic no. 121032300310-8). We thank A.A. Merdalimova for SERS measurements with LabRam HR Evolution spectrometer.

**Supporting Information.**

The Supporting Information is available free of charge on the ACS Publication website at DOI: XXX.

Section S1. Additional data for AuNR and AuNR@NBT samples.

Section S1.1. Additional TEM images (Figure S1).

Section S1.2. Histograms of the aspect ratio (AR), length (L), and diameter (D) distributions for AuNR samples 1-12 (Figures S2-S4).

Section S1.3 T-matrix simulated extinction spectra for polydisperse AuNR ensembles (Figure S5).

Section S1.4 COMSOL-simulated extinction spectra for monodisperse bare and CTAB-coated AuNRs (Figure S6).

Section S1.5 SERS measurements for AuNR@NBT and AuNR@Cy7.5 samples at 785- and 633-nm laser excitation (SERS spectra, Figures S7-S9). Figure S10. Dependence of SERS peak intensity $I_{1343}$ of AuNR@NBT conjugates at 633-nm laser excitation as a function of LPR wavelength.

Section S2. Additional data for AuNT and AuNT@NBT samples.

Section S2.1 Additional TEM images of initial and etched AuNT nanoparticles for samples 1-8 (Figures S11, S12).

S2.2 A geometrical model to characterize the shape of initial AuNT particles (Figure S13 and analysis, Eqs. S1-S6).

S2.3 COMSOL simulation of extinction spectra for a fixed AuNT orientation and comparison with experiment (Figure S14; Table S1).

Section S2.4 COMSOL simulation of extinction spectra for randomly oriented AuNT ensembles (Figures S15, S16; Eqs. S7-S11; Table S2).

Section S2.5 COMSOL simulation of extinction spectra for a three-fraction model of AuNT colloids Table S3; Figure S17).

Section S2.6 SERS spectra for AuNT@NBT and AuNT@Cy7.5 samples at 785-nm laser excitation (Figures S18, S19).

Section S2.7 Calculations of the surface and orientation averaged SERS enhancement factor for AuNR (Eqs. S12-S26; Figure S20).

Section S2.8 Morphological parameters of AuNSTs (Figures S21-23).

Section S2.9 SERS spectra for AuNST@NBT and AuNST@Cy7.5 samples at 785 nm and 633 nm laser excitation (Figures S24-S26).

**Author Contributions**

BNK: Methodology, NP synthesis, measurements, funding acquisition; AMB: TEM measurements and data processing; SVZ: COMSOL simulations; NGK: Conceptualization, methodology, T-matrix simulations, data processing, draft writing, and final editing. All authors have approved the final version of the manuscript.

**TOC Graphic**

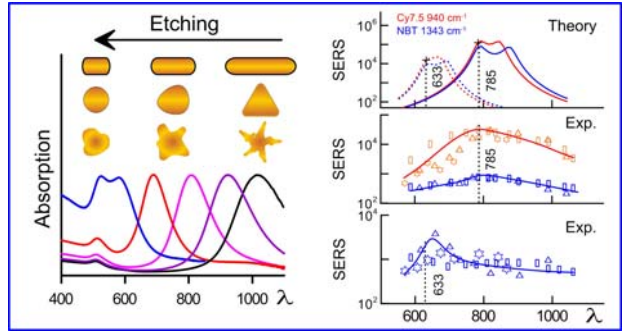

# Supporting Information

# Surface-Enhanced Raman Scattering from Au Nanorods, Nanotriangles, and Nanostars with Tuned Plasmon Resonances

*Boris N. Khlebtsov,[1] Andrey M. Burov,[1] Sergey V. Zarkov,[1] Nikolai G. Khlebtsov,[1,2,*]*

[1]Institute of Biochemistry and Physiology of Plants and Microorganisms, "Saratov Scientific Centre of the Russian Academy of Sciences," 13 Prospekt Entuziastov, Saratov 410049, Russia

[2]Saratov State University, 83 Ulitsa Astrakhanskaya, Saratov 410012, Russia

**Table of Content**

Section S2.4. COMSOL simulation of extinction spectra for randomly oriented AuNT ensembles (Figures S15, S16; Eqs. S7-S11; Table S2).

Section S2.5. COMSOL simulation of extinction spectra for a three-fraction model of AuNT colloids Table S3; Figure S17).

Section S2.6. SERS spectra for AuNT@NBT and AuNT@Cy7.5 samples at 785-nm laser excitation (Figures S18, S19).

Section S2.7. Calculations of the surface and orientation averaged SERS enhancement factor for AuNR (Eqs. S12-S26; Figure S20).

Section S2.8. Morphological parameters of AuNSTs (Figures S21-23).

Section S2.9. SERS spectra for AuNST@NBT and AuNST@Cy7.5 samples at 785 nm and 633 nm laser excitation (Figures S24-S26).

References.

# Section S1. Additional data for AuNR and AuNR@NBT samples

## Section S1.1. Additional TEM images (Figure S1)

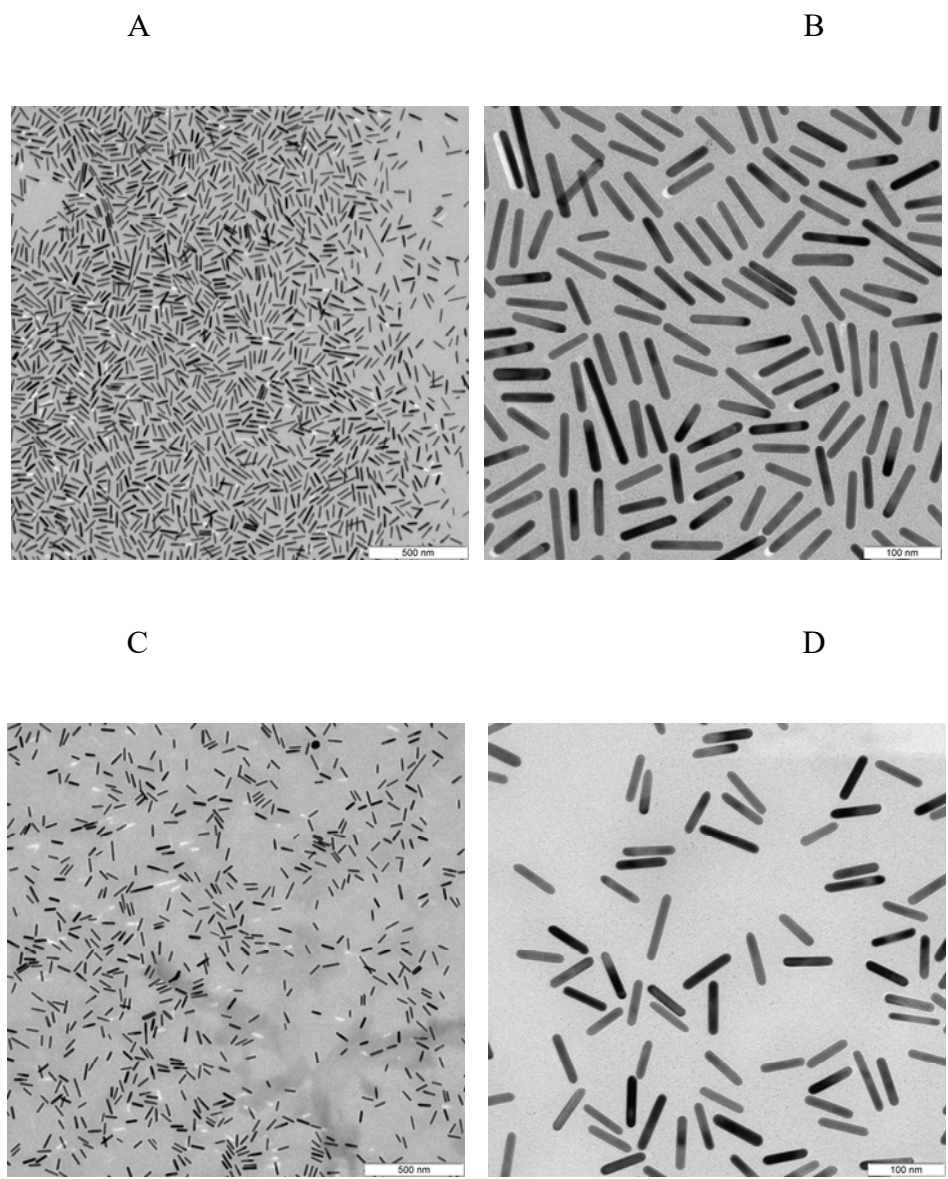

E F

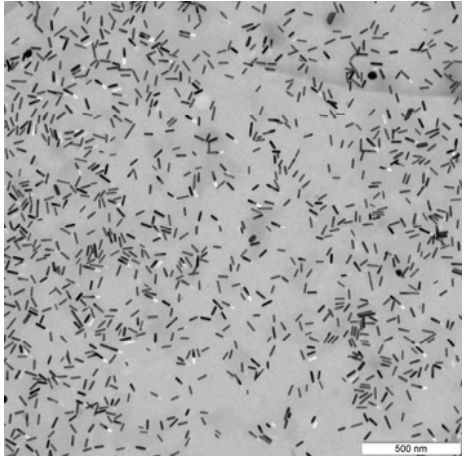 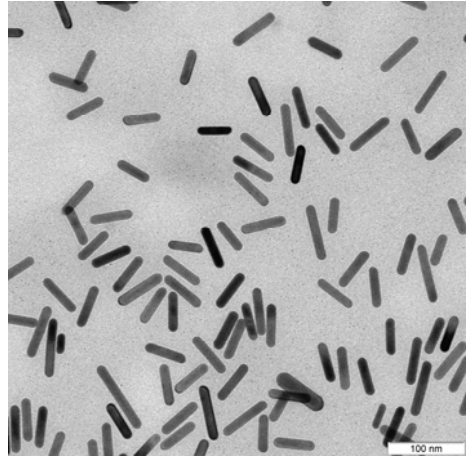

G H

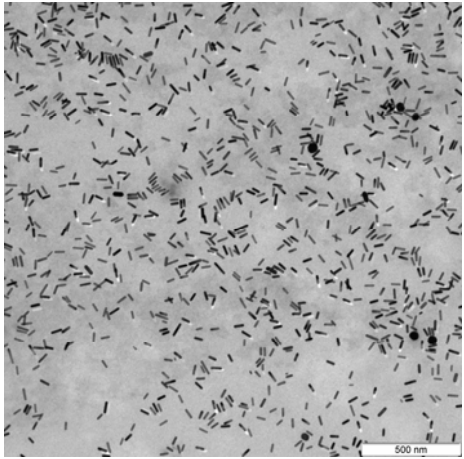 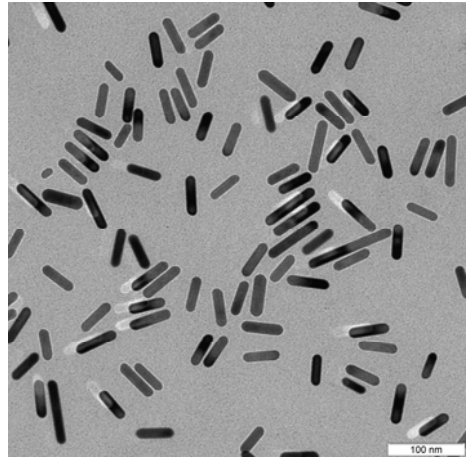

I J

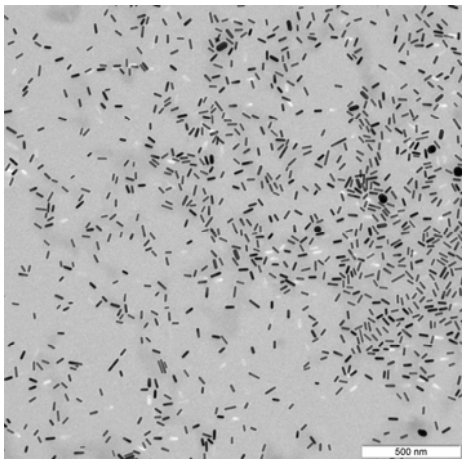 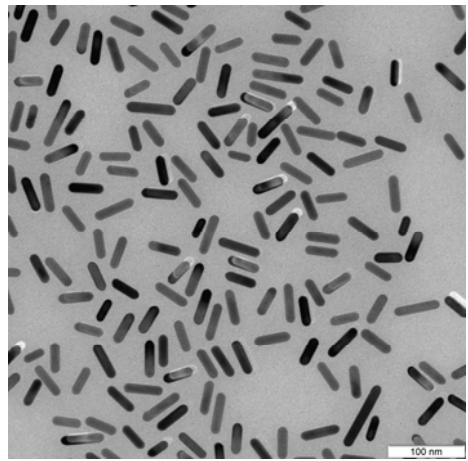

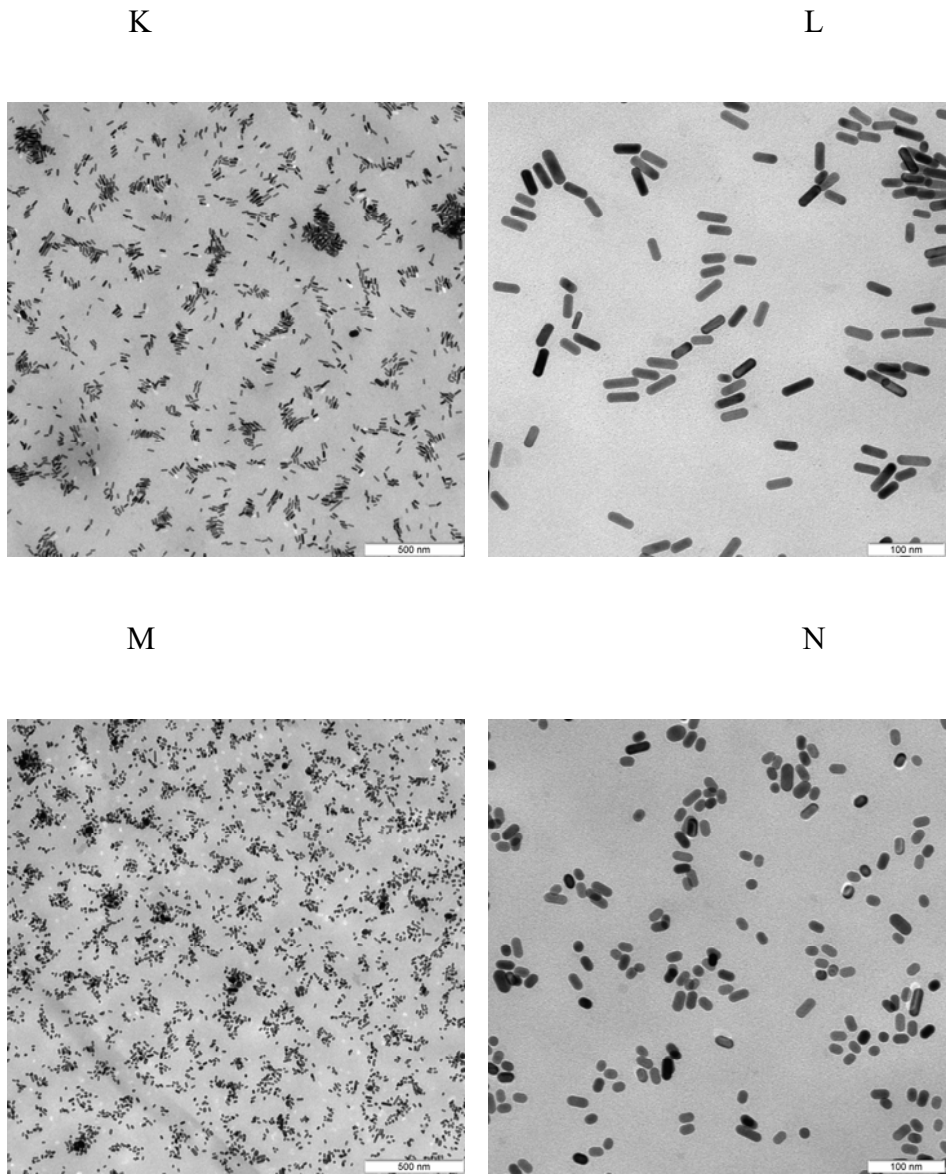

**Figure S1**. TEM images of samples 1 (A, B), 3 (C, D), 5 (E, F), 7 (G, H), 9 (I, J), 11 (K, L), and 13 (M, N). The scale bars are 500 nm (left column) and 100 nm (right column).

**Section S1.2. Histograms of the aspect ratio (AR), length (L), and diameter (D) distributions for AuNR samples 1-12.**

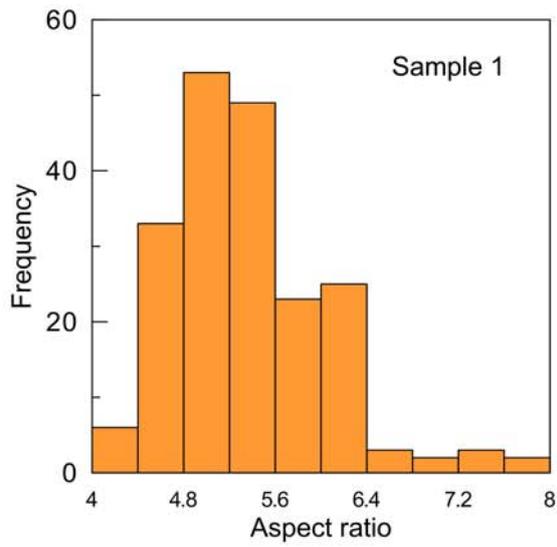
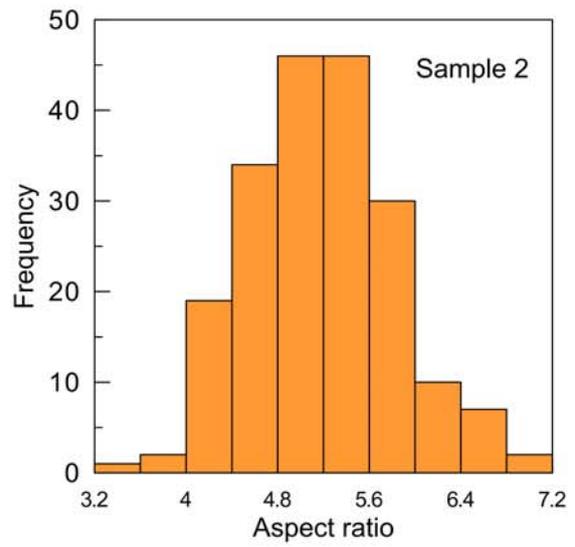
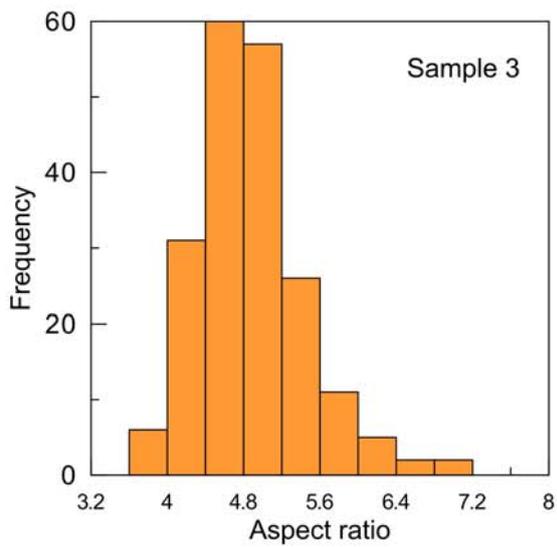
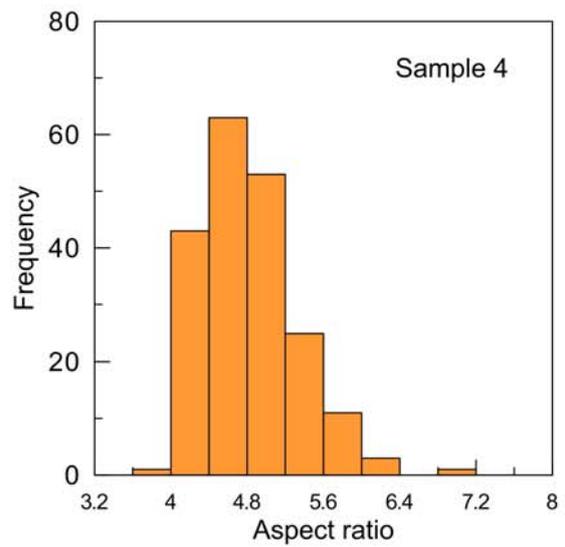
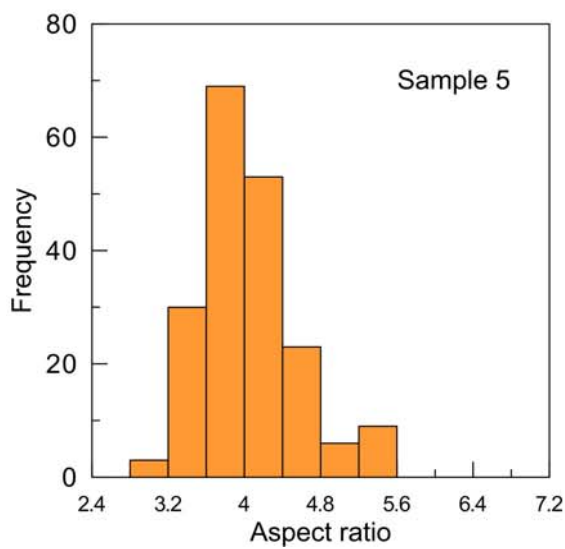
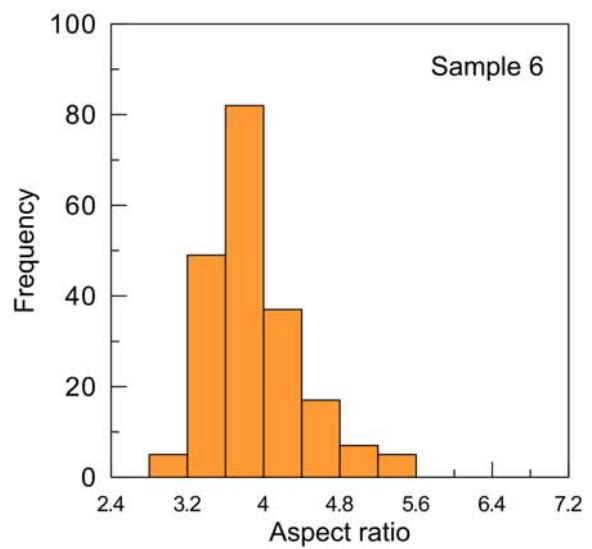

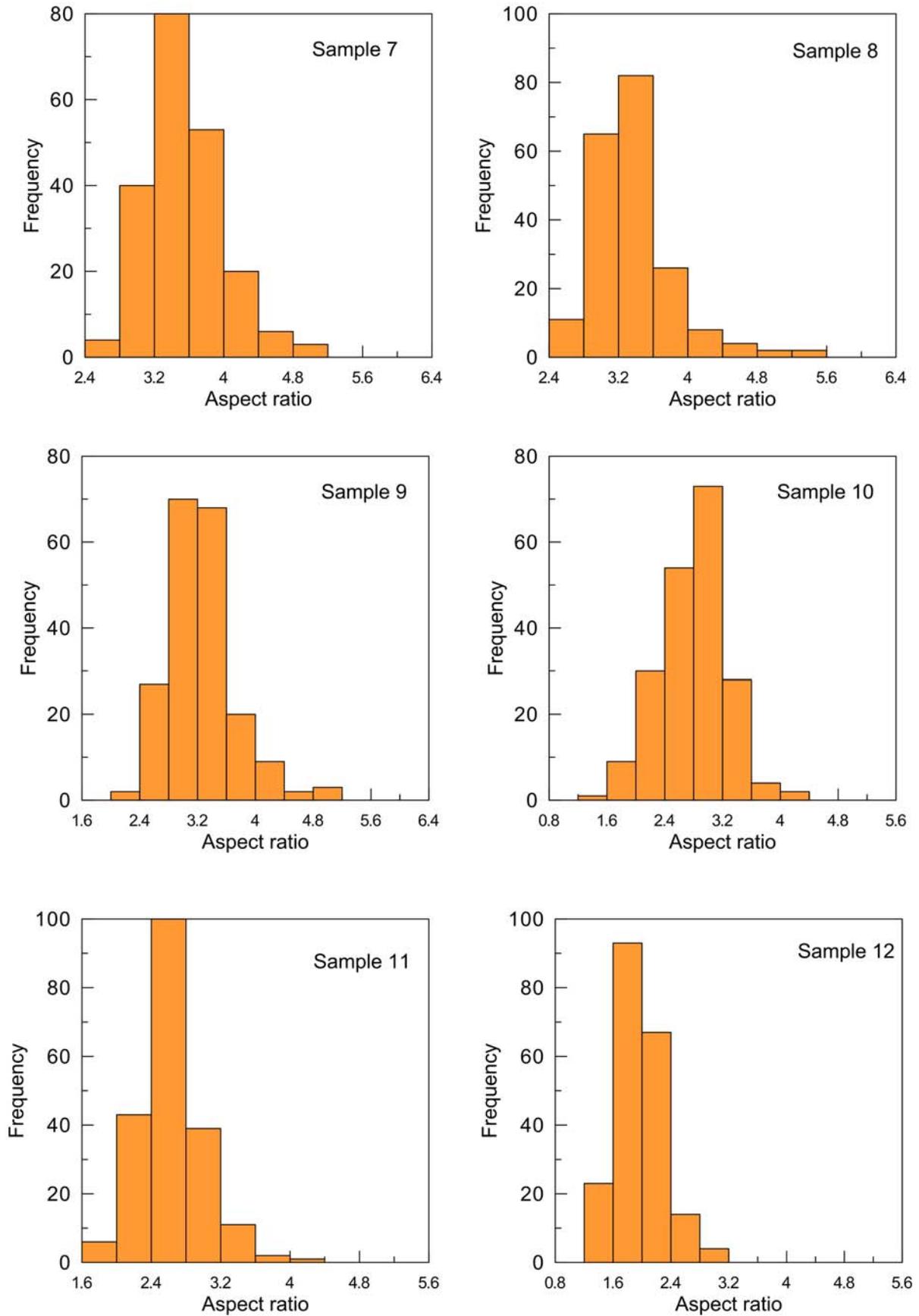

**Figure S2**. Histograms of the aspect ratio distribution for samples 1-12.

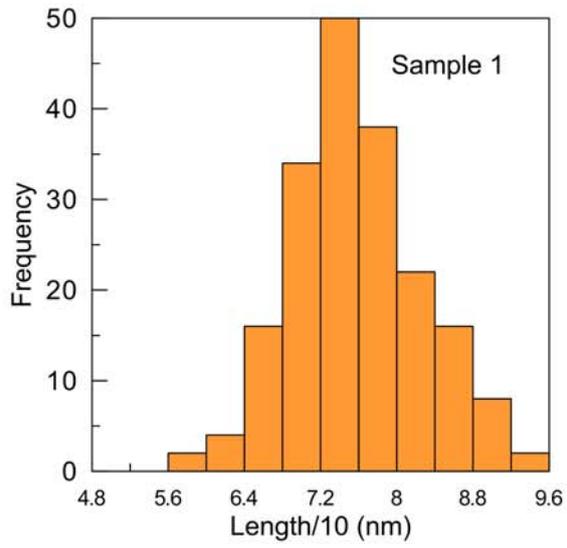
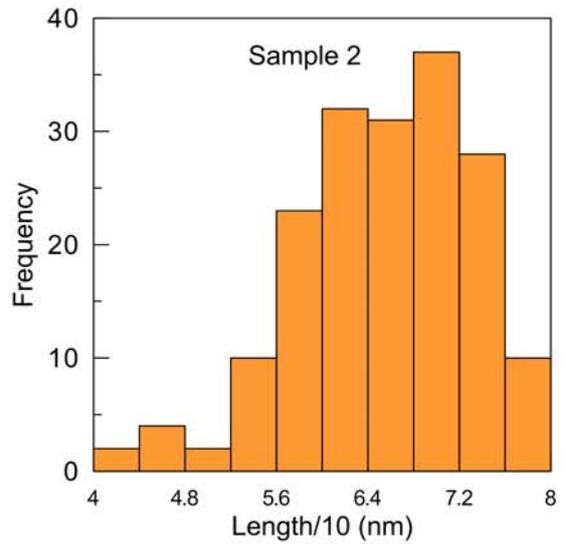
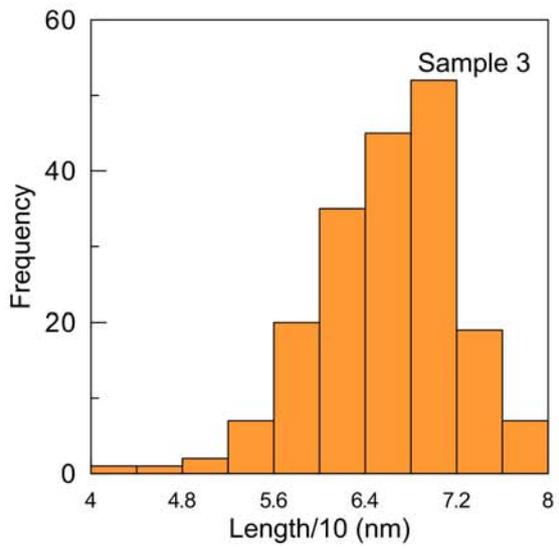
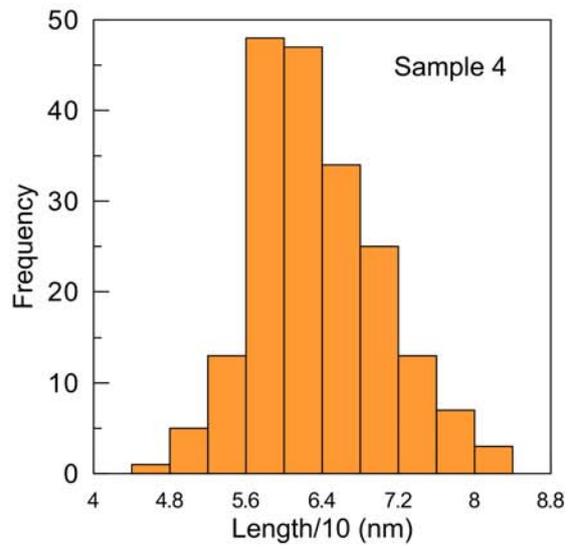
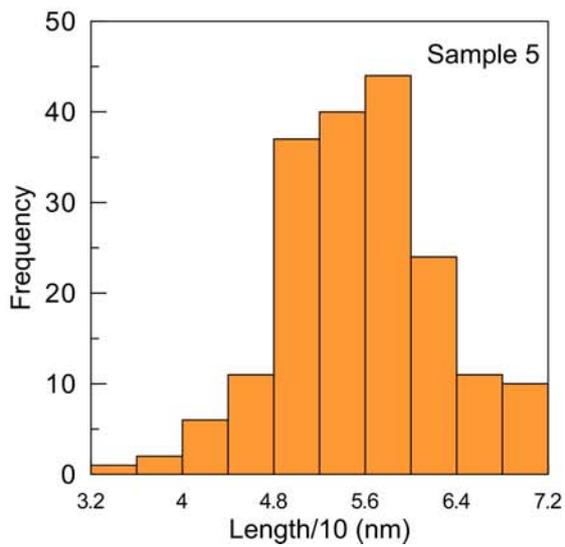
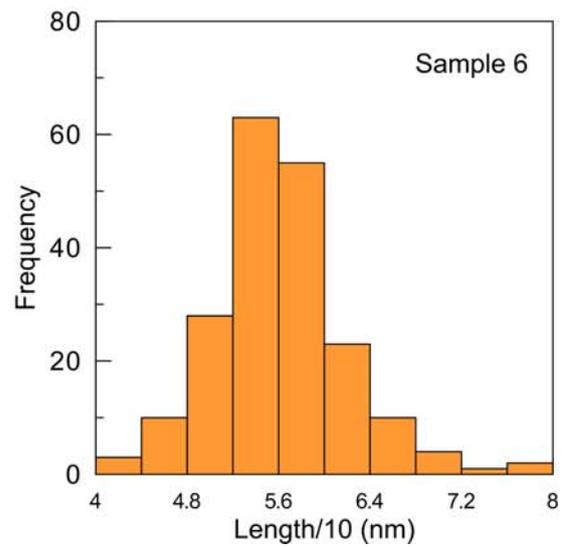

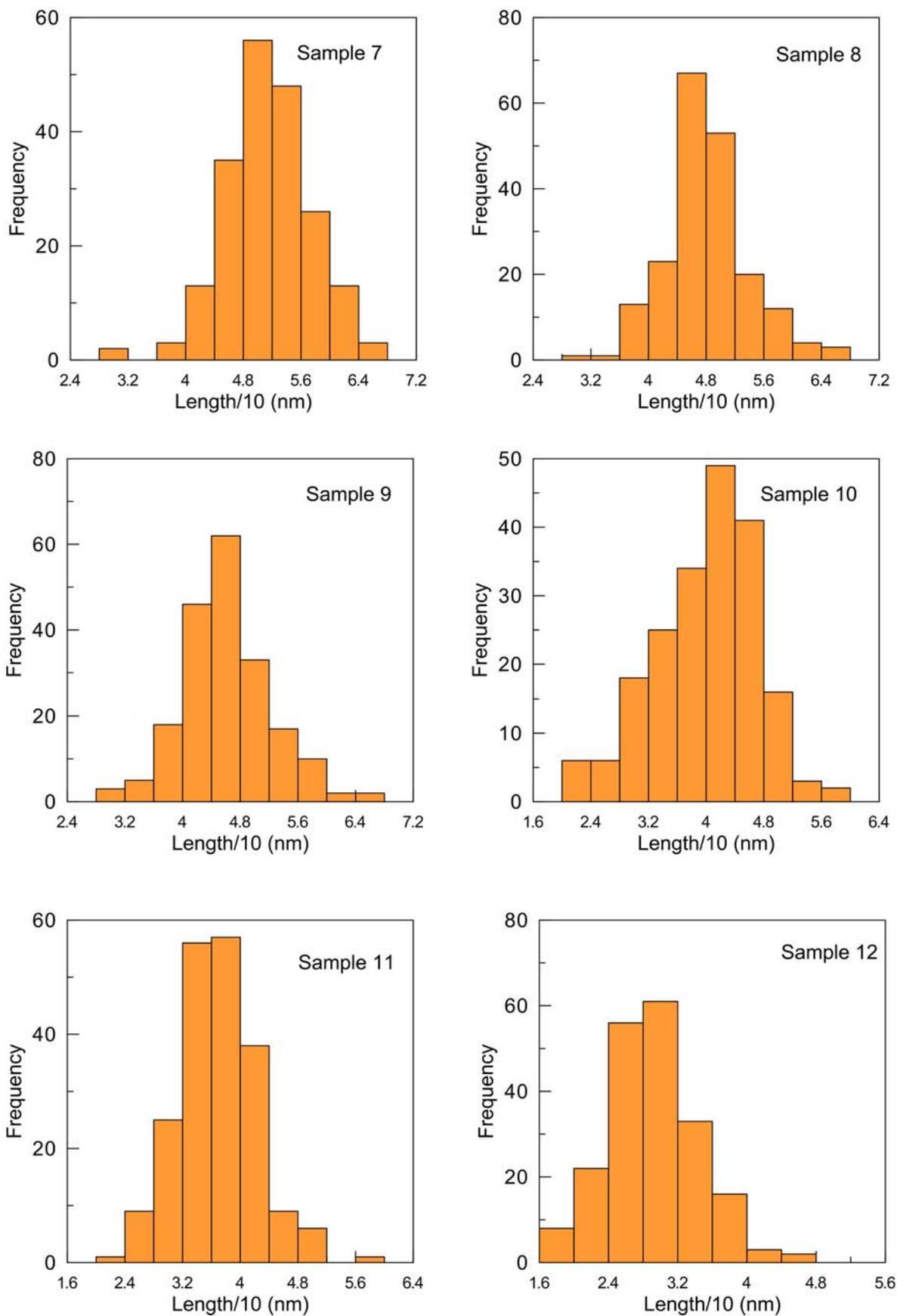

**Figure S3**. Histograms of the AuNR length distribution for samples 1-12.

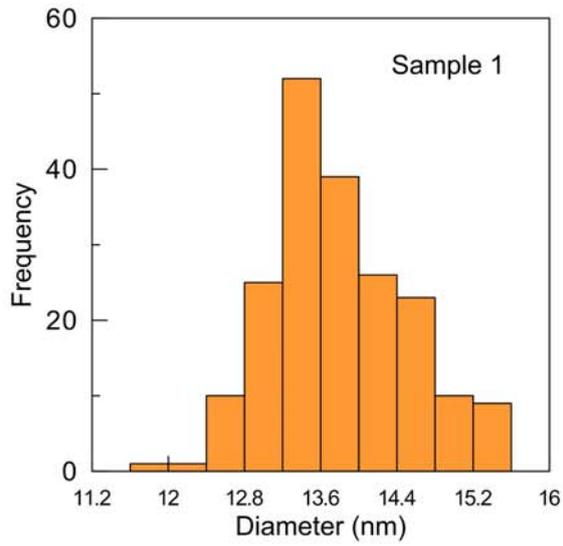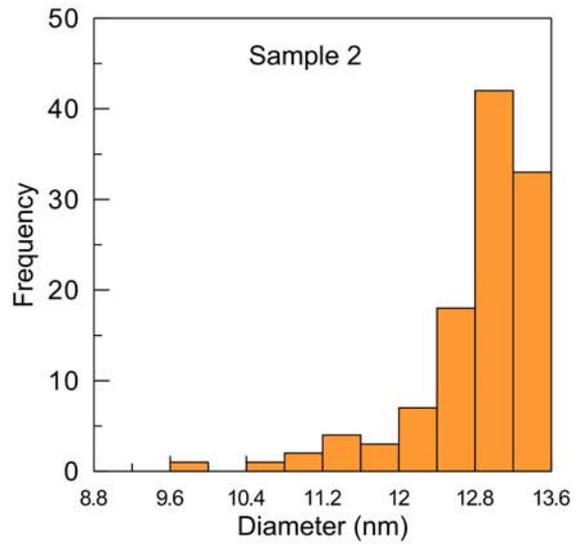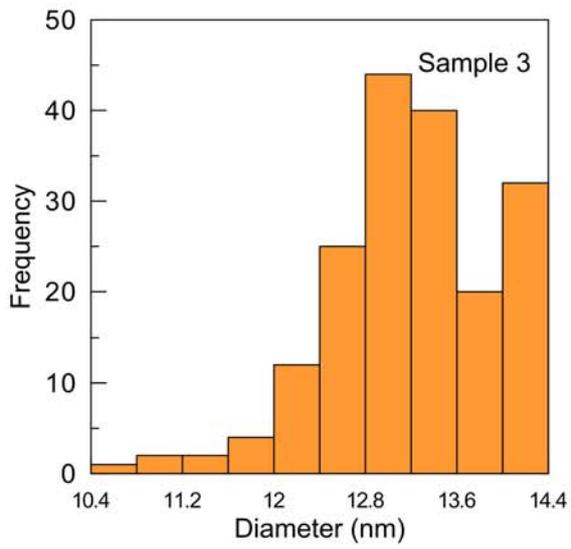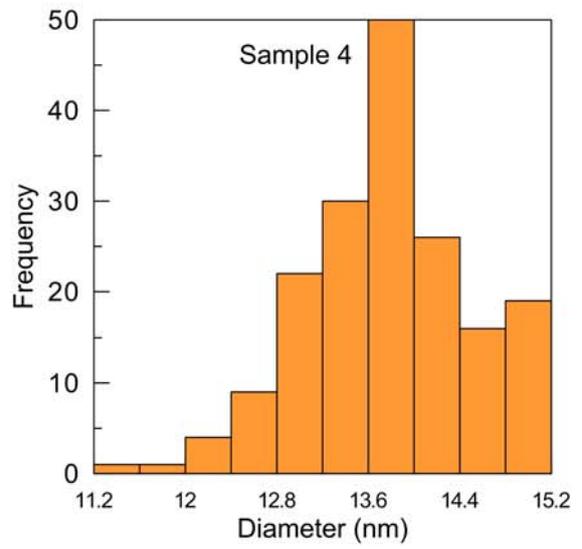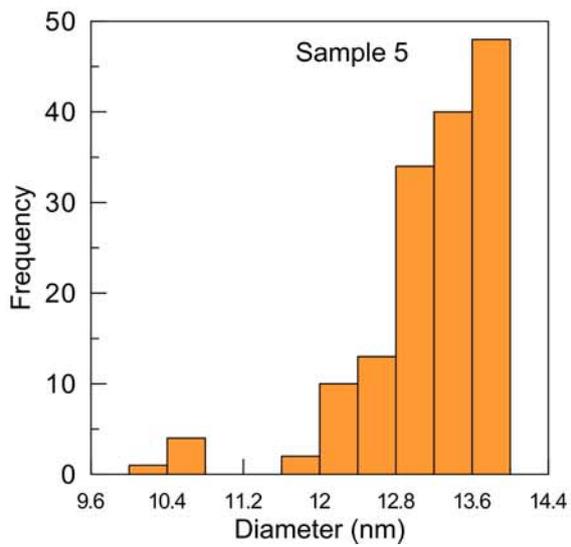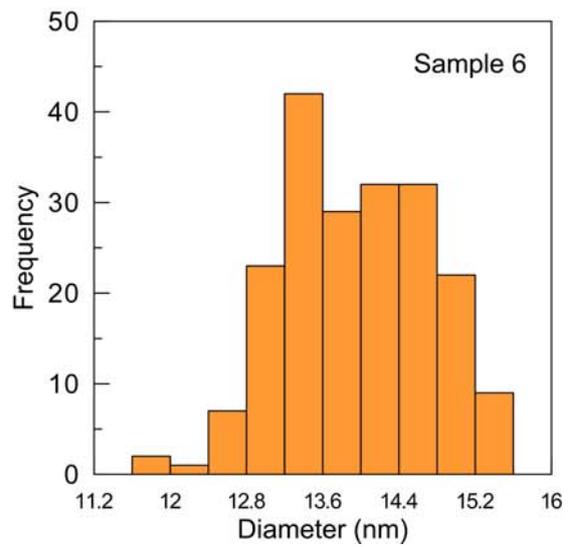

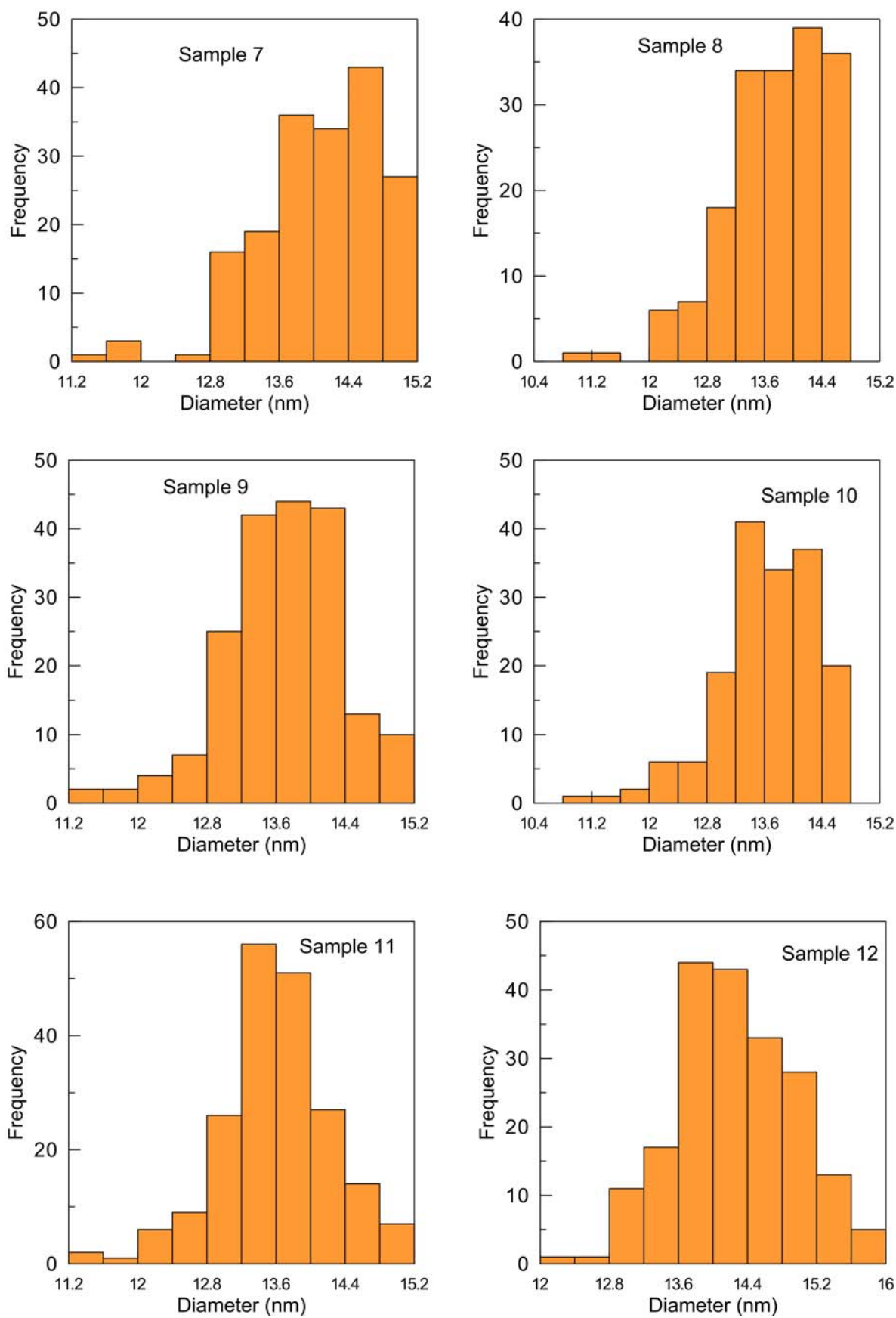

**Figure S4**. Histograms of the AuNR diameter distribution for samples 1-12.

**Section S1.3. T-matrix simulated extinction spectra for polydisperse AuNR ensembles.**

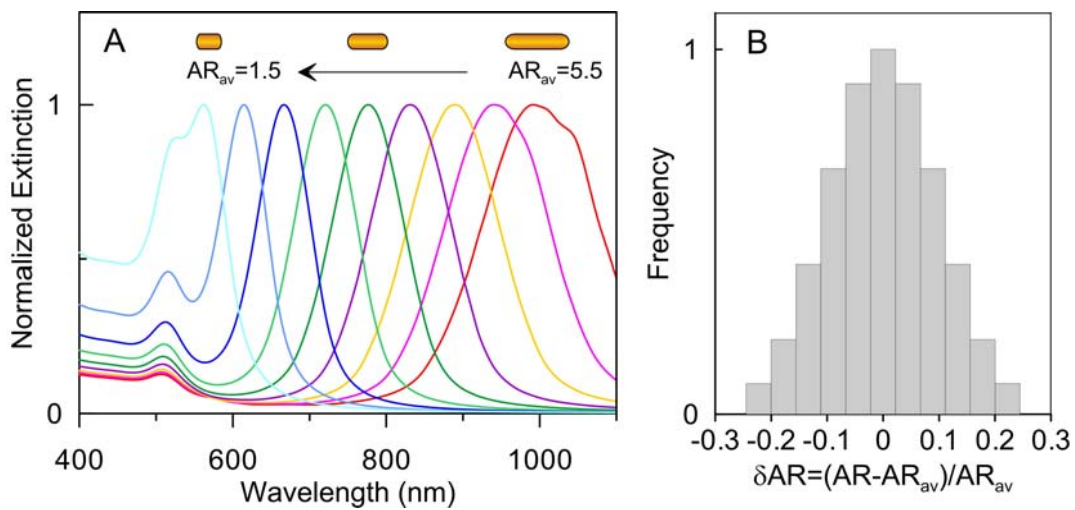

**Figure S5**. T-matrix simulated extinction spectra for polydisperse randomly oriented AuNRs with the average aspect ratios $AR_{av} = 1.5 - 5.5(0.5)$ (A). Panel B shows the aspect ratio distribution of each ensemble with relative STD =0.1 (in terms of the normalized dispersion of the Gaussian distribution). The constant average diameter equals 13.8 nm for all samples.

**Section S1.4 COMSOL-simulated extinction spectra for monodisperse bare and CTAB-coated AuNRs.**

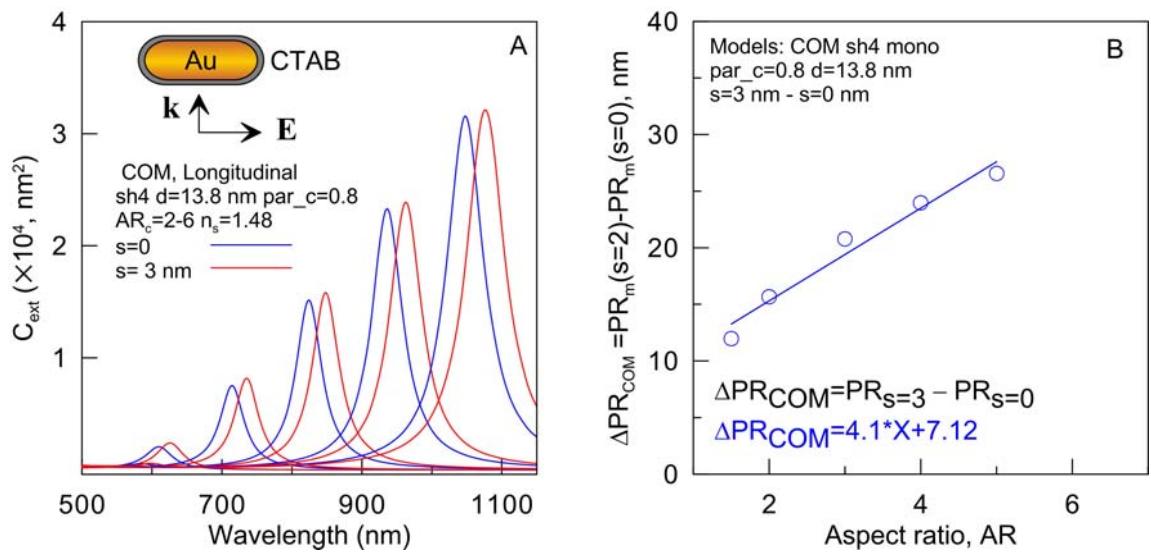

**Figure S6**. (A) COMSOL simulated extinction spectra of bare (blue) and CTAB-coated (red) AuNRs at the longitudinal excitation. The end of rods is assumed to be slightly elliptical with the cap parameter $\chi_c = b_c/(d/2) = 0.8$ ($b_c$ is the cap height, $d$ is the rod diameter).[1] Panel B shows the LPR peak shift caused by CTAB coating as a function of the aspect ratio and a linear fit. The CTAB coating is modeled by a dielectric shell of a thickness of 3 nm[2] and a refractive index of 1.48 (see the main text).

**Section S1.5 SERS measurements for AuNR@NBT and AuNR@Cy7.5 samples at 785- and 633-nm laser excitation.**

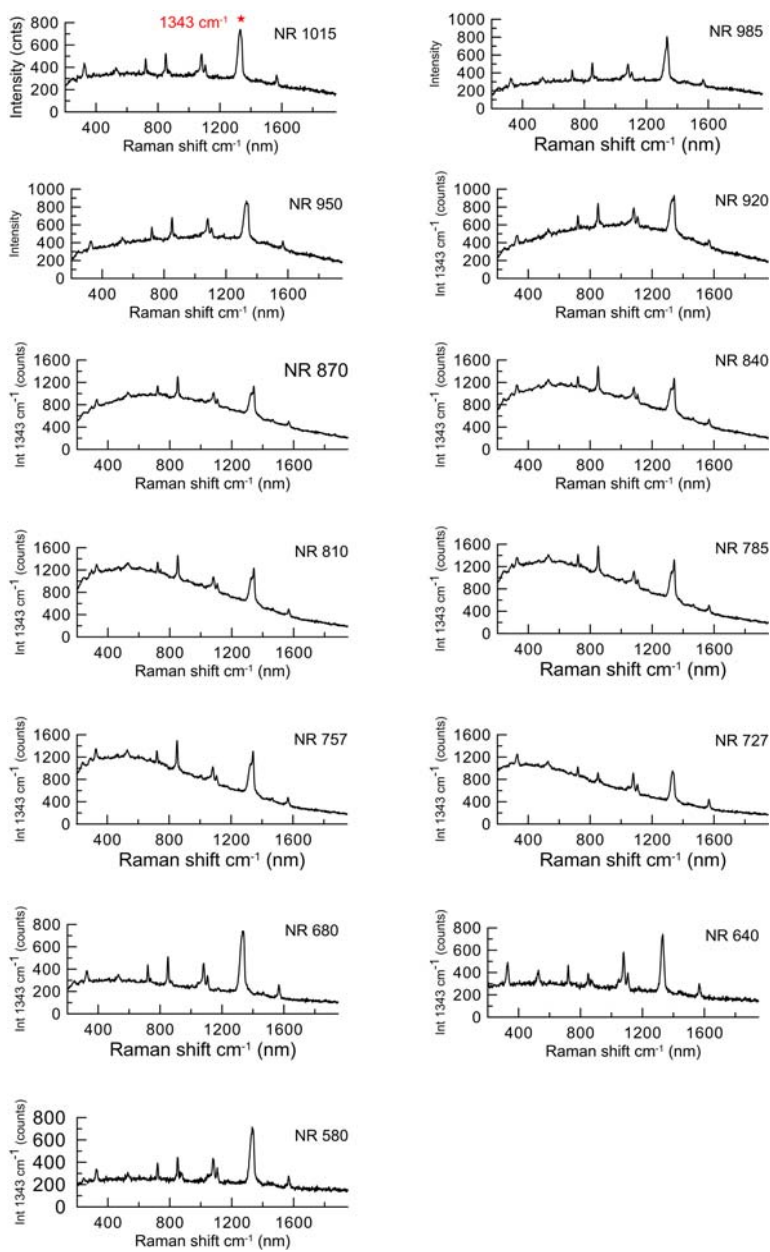

**Figure S7**. SERS spectra for 13 AuNR@NBT samples measured at 785 nm laser excitation. The LPR wavelengths varied from 1017 to 580 nm.

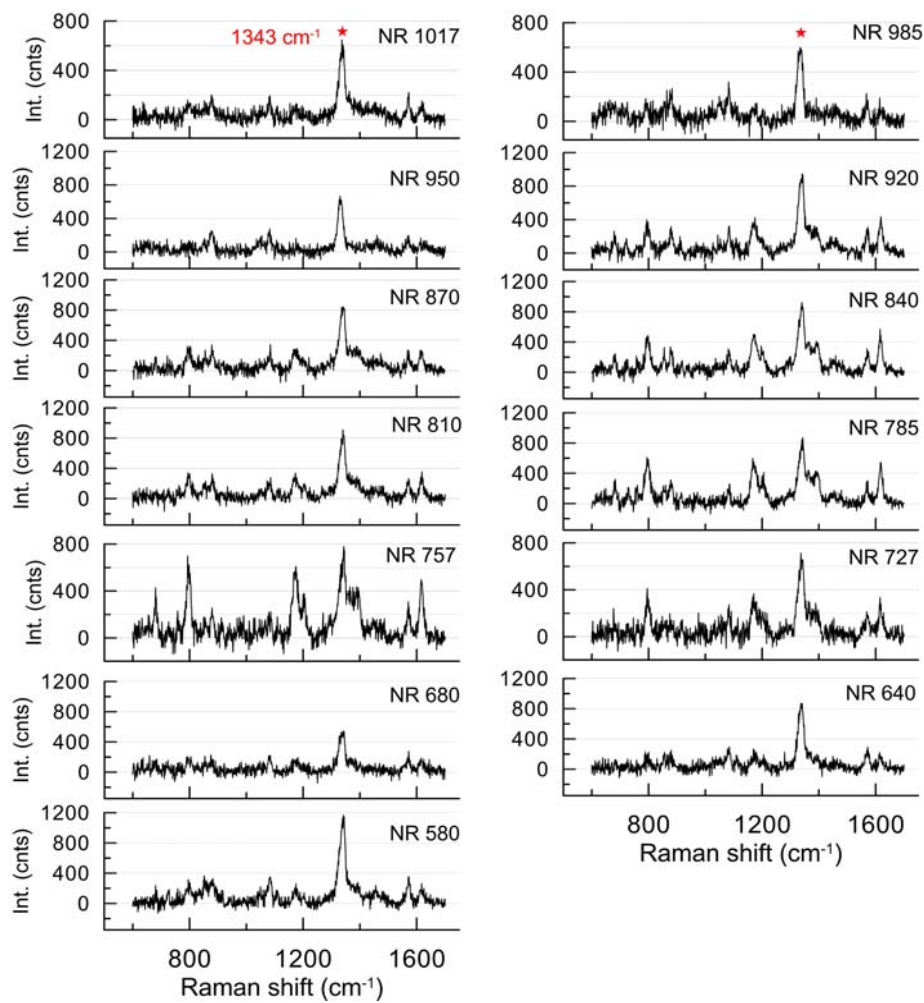

**Figure S8.** SERS spectra for AuNR@NBT samples measured at 633 nm laser excitation. The LPR wavelengths varied from 1017 to 580 nm.

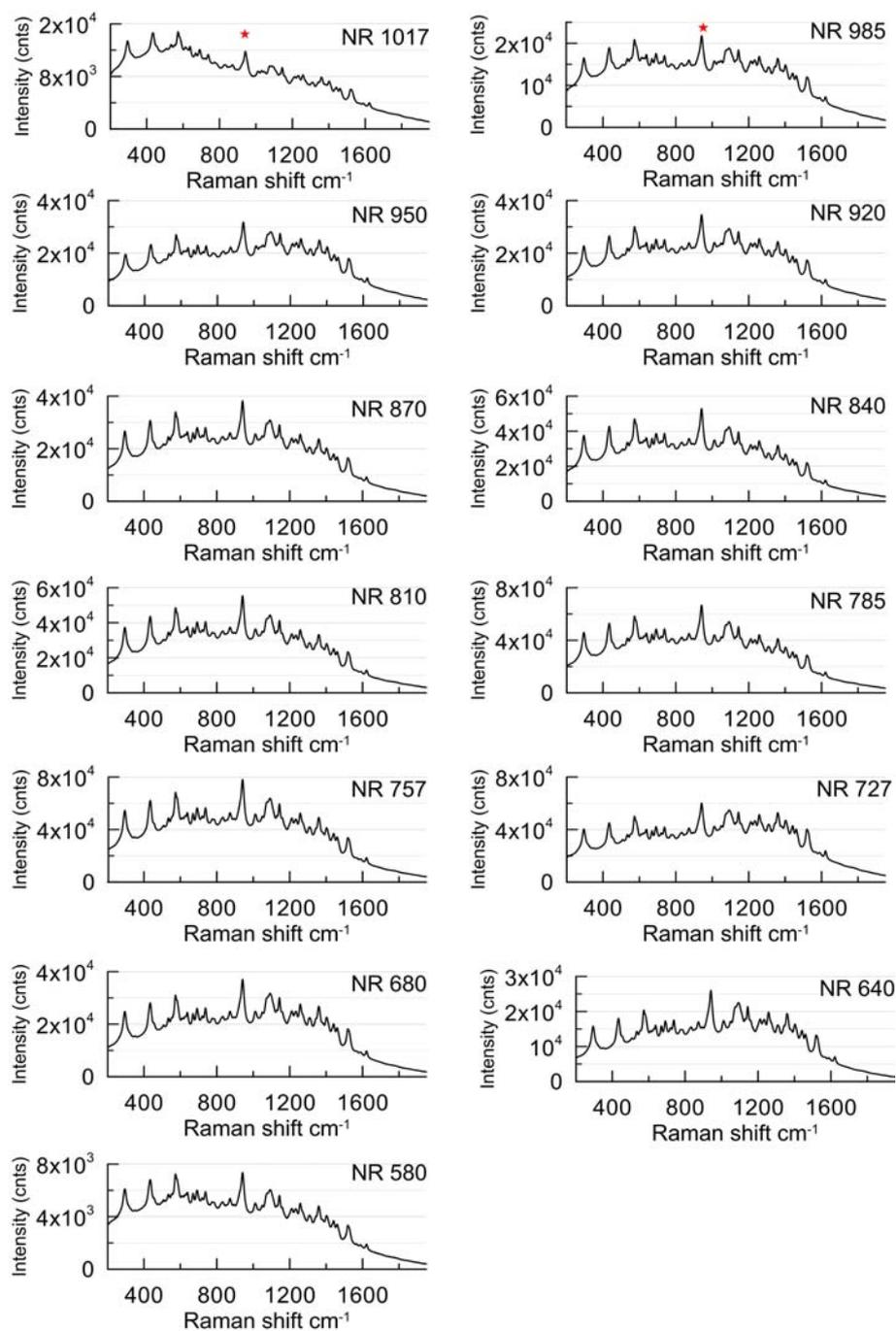

**Figure S9.** SERS spectra for AuNR@Cy7.5 samples measured at 785 nm laser excitation. The LPR wavelength decreases from 1017 to 580 nm.

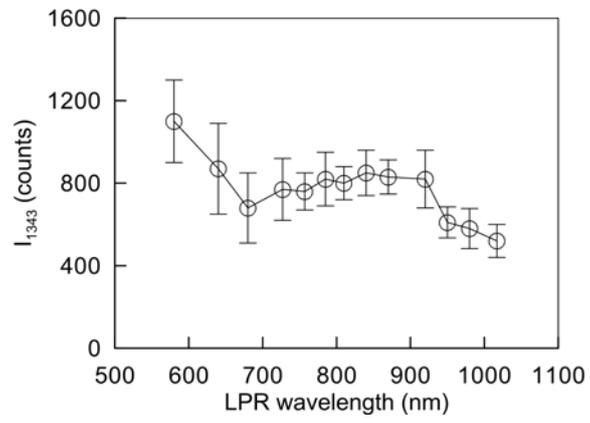

**Figure S10**. SERS peak intensity $I_{1343}$ of AuNR@NBT conjugates at 633 nm laser excitation as a function of LPR wavelength.

**Section S2. Additional data for AuNT and AuNT@NBT samples**

**Section S2.1. Additional TEM images of initial and etched AuNT nanoparticles for samples 1-8**

A

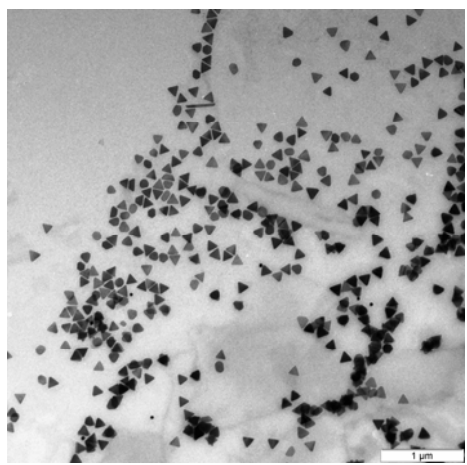

B

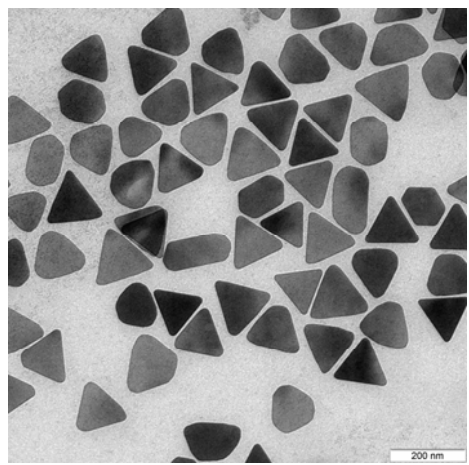

C

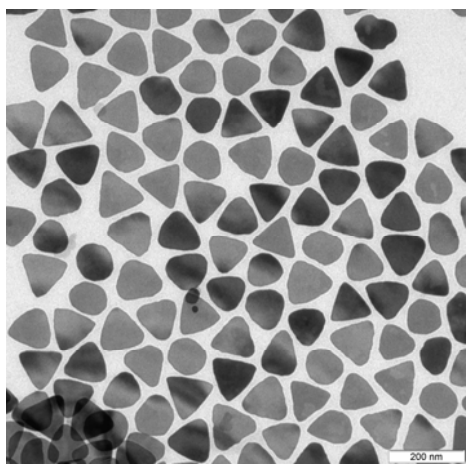

D

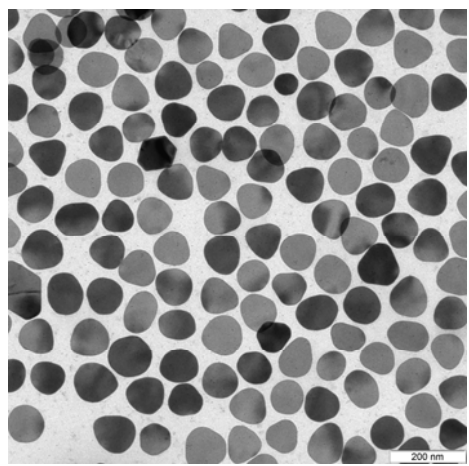

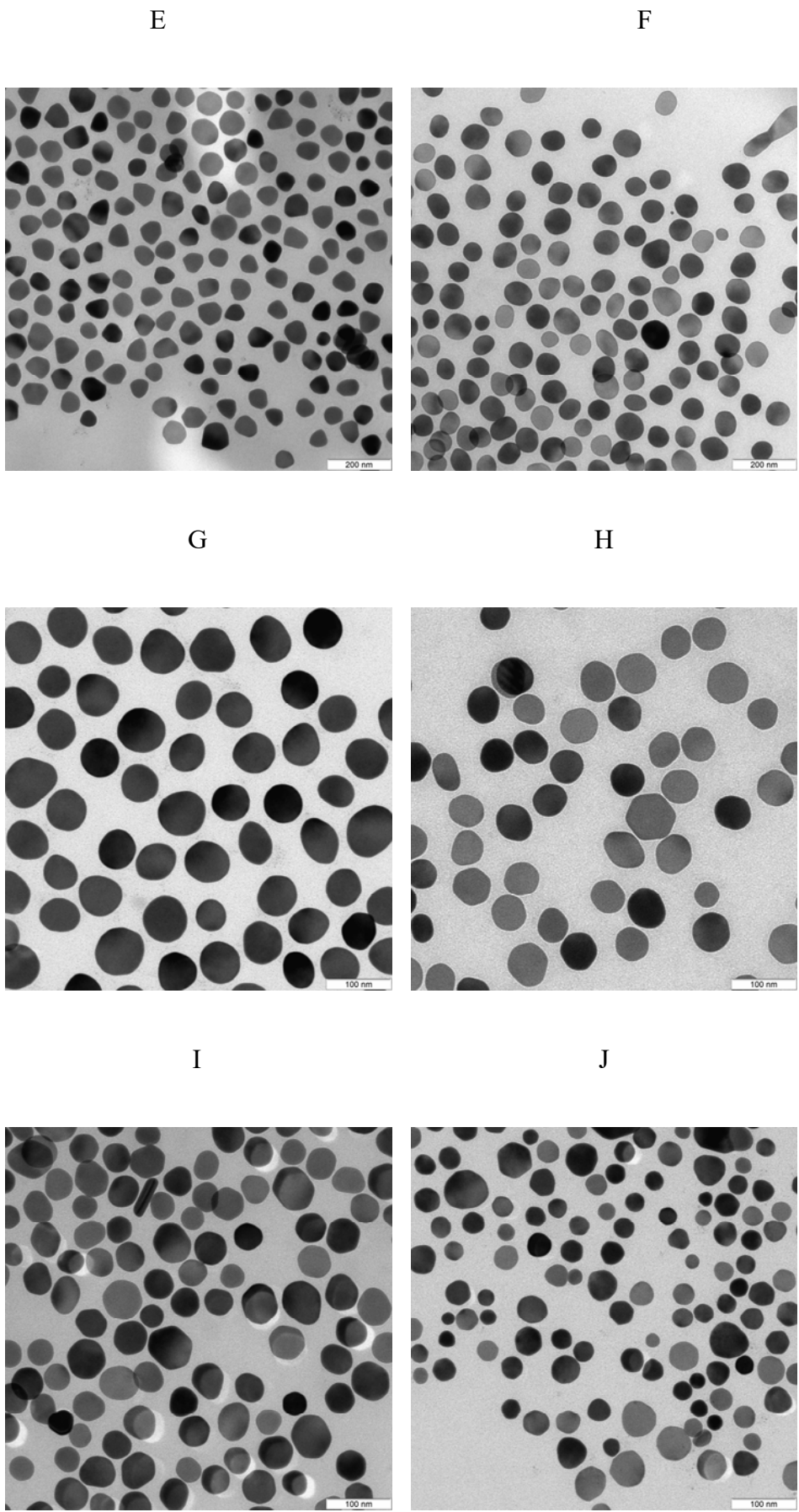

**Figure S11**. TEM images of AuNTs for the initial sample 0 (A, B) and etched samples 1-8 (C-J), respectively. The scale bars are 1 μm (A), 200 nm (B-F), and 100 nm (G-J)

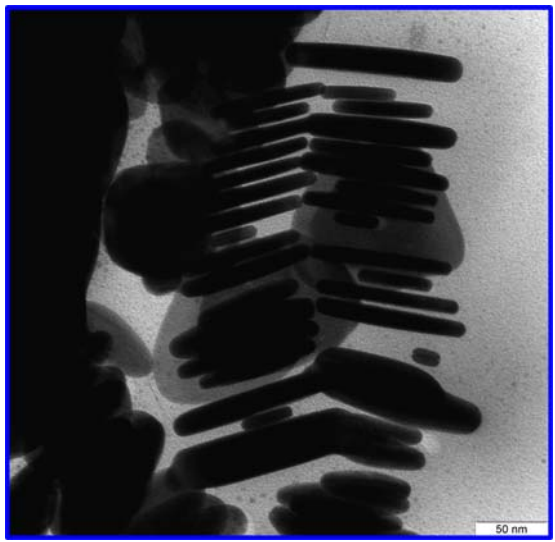

**Figure S12**. TEM image of AuNT stacks used for evaluation of AuNT thickness. The scale bar is 50 nm.

## S2.2. A geometrical model to characterize the shape of initial AuNT particles

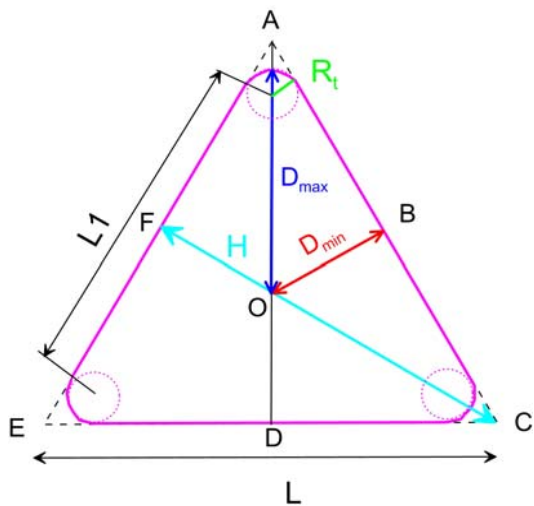

**Figure S13**. Two parameters $D_{min}$ and $D_{max}$ characterize the size and shape of Au nanotriangles. The other three parameters $L, L_1, R_t$ can be expressed through $D_{min}$ and $D_{max}$, see below, Eqs (S1-S5).

From simple geometrical considerations, one can easily derive the following relations:

$$L = 2\sqrt{3} D_{min}, \tag{S1}$$

$$L_1 = L - 2\sqrt{3} R_t, \tag{S2}$$

$$R_t = 2D_{min} - D_{max}, \tag{S3}$$

$$H = 3D_{min}. \tag{S4}$$

Assuming the AuNT area to be equal to that of the triangle AEC, we get the surface equivalent diameter

$$D_{SE} = L\sqrt{\frac{\sqrt{3}}{\pi}} = 2D_{min}\sqrt{3}\sqrt{\frac{\sqrt{3}}{\pi}} \simeq 2.57 D_{min}. \tag{S5}$$

To take into account the rounded triangle vertex, Eq. (S5) should be modified slightly

$$D_{SE}(\text{nm}) = 2.57 D_{min}(\text{nm}) - 3.5. \tag{S6}$$

In particular, for the AuNT sample 1, TEM analysis gives $D_{min} = 43.4$ nm and $D_{SE} = 108.1$ nm. From Eq. (S5), we have almost the same value $D_{SE} = 108$ nm.

**Section S2.3. COMSOL simulation of extinction spectra for a fixed AuNT orientation and comparison with experiment.**

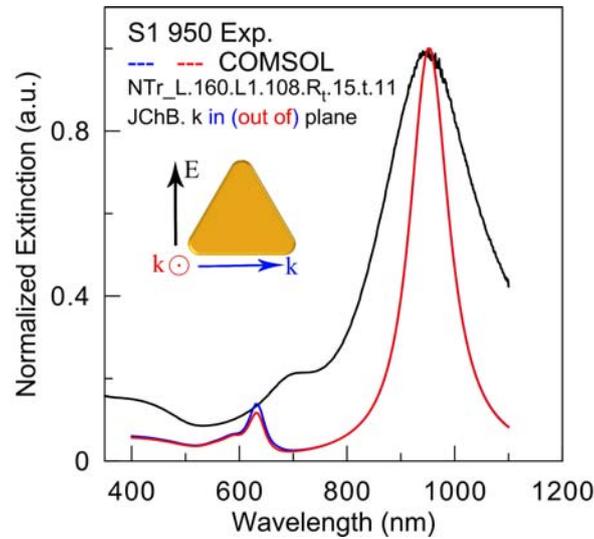

**Figure S14**. Comparison of experimental (black) and COMSOL-simulated (blue and red) extinction spectra. The simulations were performed for an AuNT model with the following parameters: $L = 160$ nm, $R_t = 15$ nm, thickness $t = 11$ nm, the incident field **E** lies in-plane, and the wave vector **k** lies in or out-of-plane. The dielectric function was taken from the bulk Johnson-Christy data. While the experimental and simulated major peak positions agree, the positions of multipole peaks differ by 70 nm. Note very close simulated spectra for in-plane and out-of-plane incidence.

**Table S1.** Extinction cross sections calculated at three resonance wavelengths for in-plane and out-of-plane incidence and in-plane excitation (Figure S12)

| $\lambda$(nm) | $C_{ext}$(nm$^2$) | | Difference (%) |
| --- | --- | --- | --- |
| | In-plane | Out-of-plane | |
| 590 | 7904.7 | 7908.4 | -0.05 |
| 635 | 14475.7 | 16545.5 | -14.3 |
| 955 | 125163 | 119936 | 4.2 |

## Section S2.4. COMSOL simulation of extinction spectra for randomly oriented AuNT ensembles

Due to particle symmetry, we consider the scattering geometry shown in Figure S13. Without any loss of generality, the nanotriangle is placed in the (*x,z*) plane, with one side being directed along the x-axis. Angles $\theta, \varphi$ specify an arbitrary direction of the electric field.

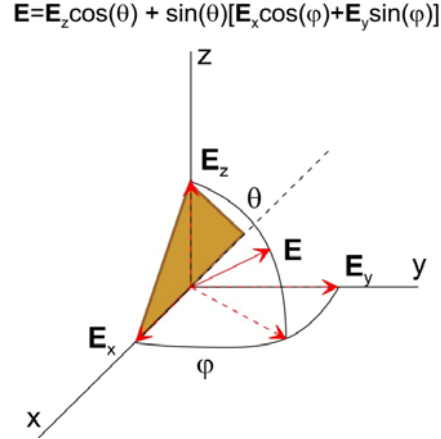

**Figure S15.** The geometry of light scattering by an Au nanotriangle.

According to our preliminary COMSOL simulations, the extinction cross sections virtually do not depend on the direction of the wave vector. Thus, we have, in fact, an electrostatic problem. This means that the averaging over the random orientation of particles is equivalent to the averaging over the electric field directions. In electrostatic approximation, the extinction cross section can be written as

$$C_{ext} = C_{ext}^{z}\cos^2(\theta) + \sin^2(\theta)\left[C_{ext}^{x}\cos^2(\varphi) + C_{ext}^{y}\sin^2(\varphi)\right]. \tag{S6}$$

After averaging Eq. S6 over angles $\varphi$, we get a standard electrostatic expression

$$C_{ext} = \frac{C_{ext}^{z} + C_{ext}^{x} + C_{ext}^{y}}{3}. \tag{S7}$$

We have to stress that Eq. (S7) is an approximation even though the cross sections for x, y, and z polarization are calculated by an exact method. Unfortunately, Eq. (S7) is sometimes considered in the literature as the right numerical recipe,[3] Taking into account that $C_{ext}^z \simeq C_{ext}^x$ (Figure S12), we finally get

$$C_{ext} = \frac{2C_{ext}^z + C_{ext}^y}{3}. \tag{S8}$$

Equation (5) assumes that the cross section $C_{ext}^z$ not only coincides with $C_{ext}^x$ but does not also depend on the direction of the wave vector. In fact, calculations have shown that there are differences in values $C_{ext}^z$ for the incidence of light perpendicular to the plane of the particle (i.e., along the y-axis) and on the end of the particle, i.e., along the x-axis. Therefore, as we have shown, such a formula gives more accurate averaged cross sections

$$C_{ext} = \frac{C_{ext}^z(\mathbf{k}_y) + C_{ext}^z(\mathbf{k}_x) + C_{ext}^y(\mathbf{k}_z)}{3}. \tag{S9}$$

As in the general case, the angular dependence of the cross section $C_{ext}(\theta,\varphi)$ can differ from electrostatic approximation Eq. S1, the direct numerical averaging should be performed by the following integral

$$\langle C_{ext} \rangle = \int_0^1 dy \int_0^1 dx\, C_{ext}(x = \cos\theta, y = 2\varphi/\pi). \tag{S10}$$

In a further approximation, we assume that the $\varphi$ dependence of cross sections can be expressed similarly to Eq. (S6), and the particular cross sections $C_{ext}^x, C_{ext}^y$ do not depend on the azimuth. Then, we can perform the averaging of cross sections over $\varphi$ analytically to get

$$\langle C_{ext} \rangle = \int_0^1 C(x = \cos\theta, \varphi = \pi/2) dx. \tag{S11}$$

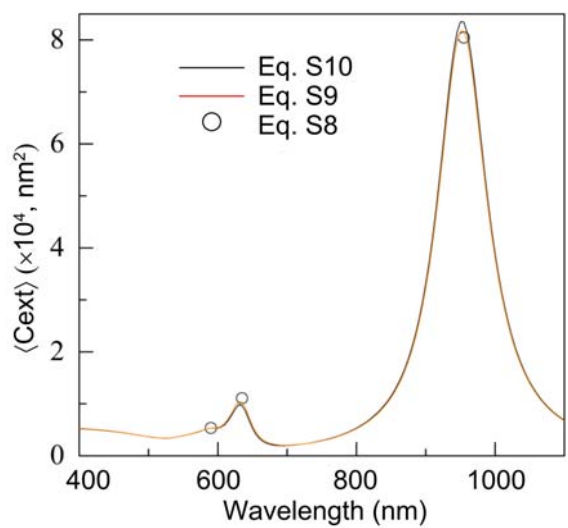

**Figure S16.** Comparison of orientation-averaged extinction cross sections calculated by Eq. S11 (black) and approximations Eq. S10 (red) and Eq. S9 (circles). Model parameters are the same as in Figure S12. Note the excellent approximation accuracy given by Eq. S10 (see also Table S2).

**Table S2.** Orientation averaged cross sections calculated for three resonance wavelengths.

| $\lambda$(nm) | $\langle C_{ext} \rangle$ (nm²) | | | Error (%) | |
|---|---|---|---|---|---|
| | Eq. S10 | Eq. S9 | Eq. S8 | Eq. S9 | Eq. S8 |
| 590 | 5353.3 | 5331.7 | 5330.5 | -0.4 | -0.4 |
| 635 | 11072.8 | 10368.4 | 9678.5 | -6.4 | -12.6 |
| 955 | 80453.5 | 81703.9 | 83446.3 | 1.6 | 3.7 |

**Section S2.5. COMSOL simulation of extinction spectra for a three-fraction model of AuNT colloids**

Examination of TEM images revealed three main particle types in AuNT colloids: AuNT triangles, elliptic discs (ellipses), and circular disks. During the etching process, the initial major fraction of nanotriangles is transformed sequentially into disks and elliptic disks. Such etching is accompanied by a gradual decrease in the average particle size in terms of the surface-equivalent diameter. Table S3 summarizes TEM-derived data on the percentage and geometrical parameters of three fractions for all 8 samples S1-S8. The particles' shape and geometrical parameters are shown in the pictures in the upper first table row. After the etching process has been completed, the major nanotriangle fraction (~80%) is transformed into the major disk fraction (72 %) and a minor elliptic disc fraction (28 %).

Using the above tree-fraction model and the orientation averaging procedure described by Eq. S8, we calculated the extinction spectra of samples S1-S8 (Figure S15). The normalized extinction spectra generally agree with the experimental plots in Figure 3F. However, the LPR peak position of sample 1 is blue-shifted by ~50 nm. Further, the FWHMs of all calculated spectra are less than the measured ones. These differences can be explained by many factors, including the approximate character of the model (Table S3) used in simulations.

Table S3. Fraction percentage and geometrical parameters of etched particles in model colloids

| Sample | Triangle | | | | Disk | | Ellipse | | |
|---|---|---|---|---|---|---|---|---|---|
| | Percentage % | l nm | h nm | r nm | Percentage % | D nm | Percentage % | L nm | H nm |
| Sample 1 | 79.3 | 64.6 | 43.0 | 12.5 | 3.3 | 100.7 | 17.4 | 116.3 | 92.0 |
| Sample 2 | 71.4 | 62.3 | 46.1 | 26.3 | 16.1 | 105.7 | 12.5 | 122.1 | 103.2 |
| Sample 3 | 15.5 | 55.6 | 42.5 | 30.7 | 57.1 | 98.4 | 27.4 | 110.0 | 89.2 |
| Sample 4 | 4.8 | 50.3 | 39.7 | 28.4 | 71.4 | 75.2 | 23.8 | 85.0 | 71.5 |
| Sample 5 | 4.2 | 40.7 | 34.4 | 26.5 | 75.0 | 75.6 | 20.8 | 82.8 | 66.4 |
| Sample 6 | 0 | | | | 66.7 | 65.4 | 33.3 | 68.8 | 56.5 |
| Sample 7 | 0 | | | | 70.3 | 56.9 | 29.7 | 62.8 | 51.3 |
| Sample 8 | 0 | | | | 72.4 | 49.1 | 27.6 | 63.0 | 52.2 |

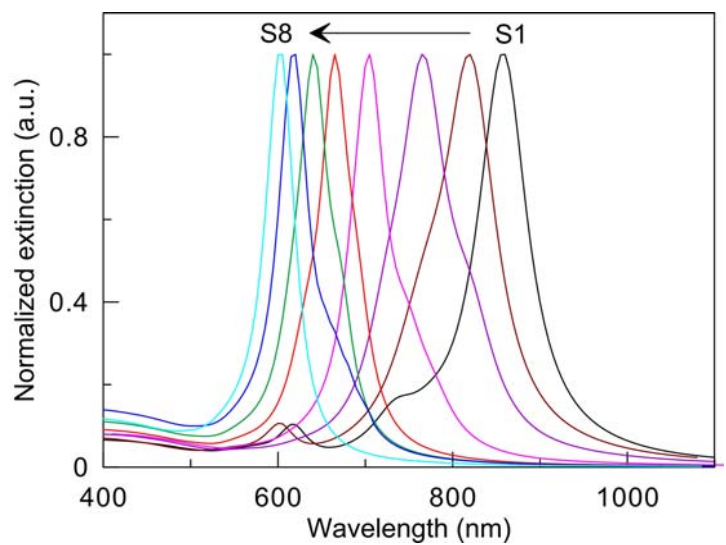

**Figure S17.** Extinction spectra were simulated for the three-fraction model using colloid compositions and particle parameters in Table S3.

**Section S2.6. SERS spectra for AuNT@NBT and AuNT@Cy7.5 samples at 785-nm laser excitation.**

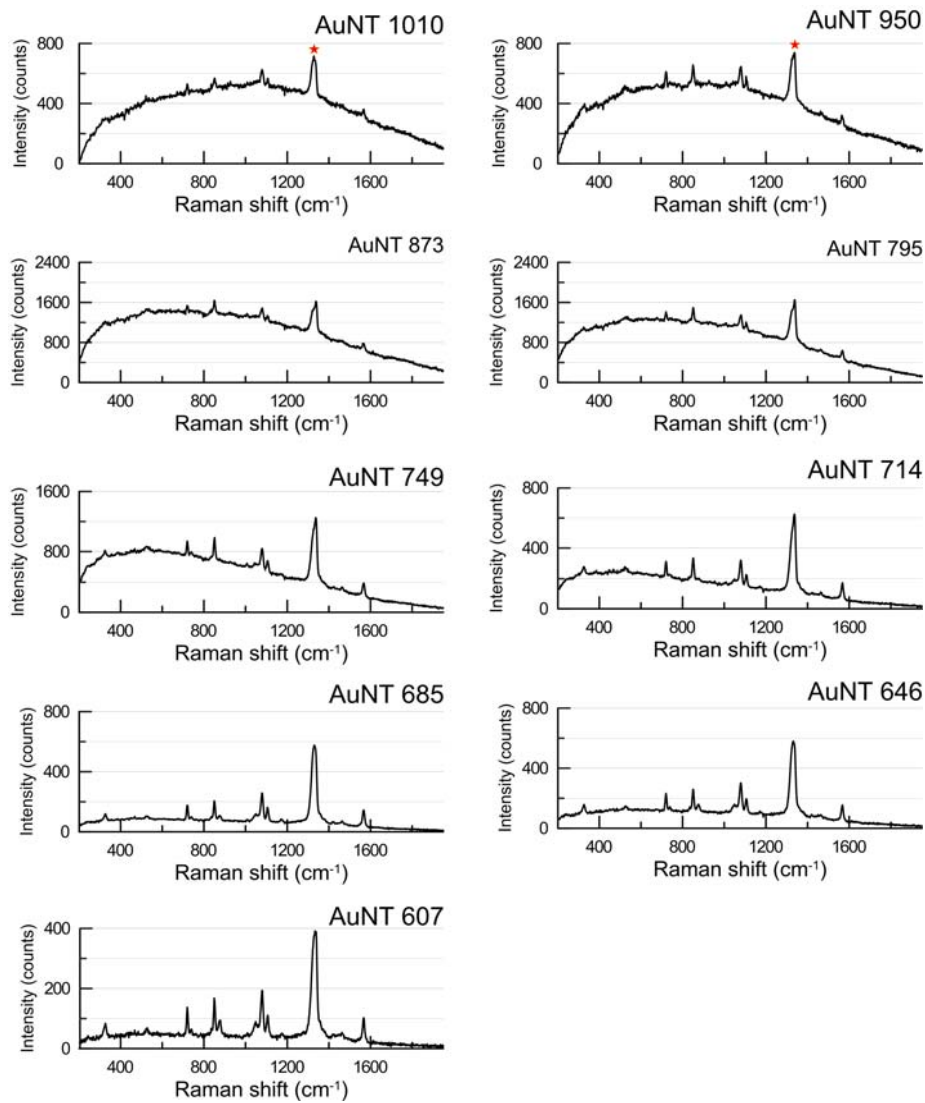

**Figure S18**. SERS spectra for AuNT@NBT samples measured at 785 nm laser excitation. The LPR wavelength decreases from 1010 to 607 nm.

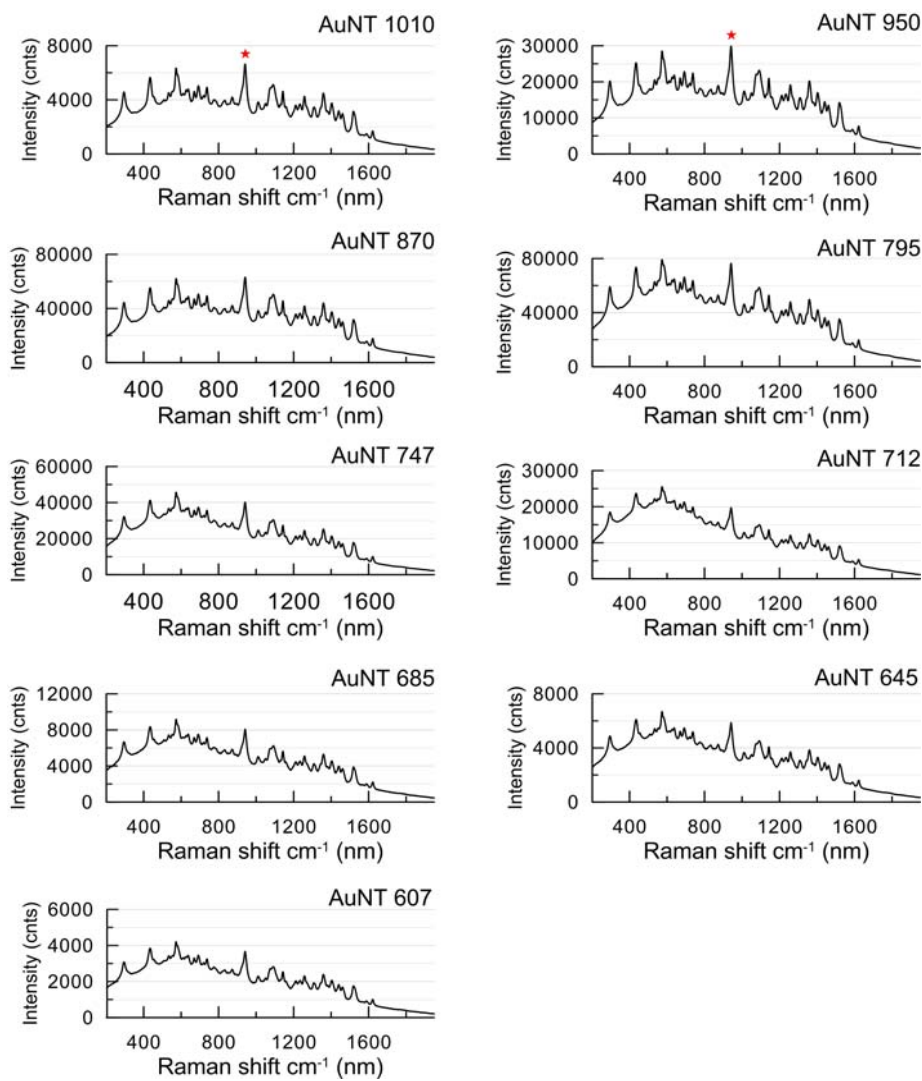

**Figure S19.** SERS spectra for AuNT@Cy7.5 samples measured at 785-nm laser excitation. The LPR wavelength decreases from 1010 to 607 nm.

**Section S2.7. Calculations of the surface and orientation averaged SERS enhancement factor for AuNRs**

Consider first the averaging of EF over the total outer AuNR surface $S$. By definition, $\langle EF \rangle_s$ equals[4]

$$\langle EF \rangle_s = \frac{1}{S} \int_S |E(\omega_L)|^2 |E(\omega_R)|^2 \, dS, \tag{S12}$$

where the squared modulus of fields are taken at the laser and Raman frequencies. Owing to the nanorod symmetry, it is sufficient to consider the upper half of the particle (Figure S20) and perform the averaging over the polar angles $\theta, \vartheta$.

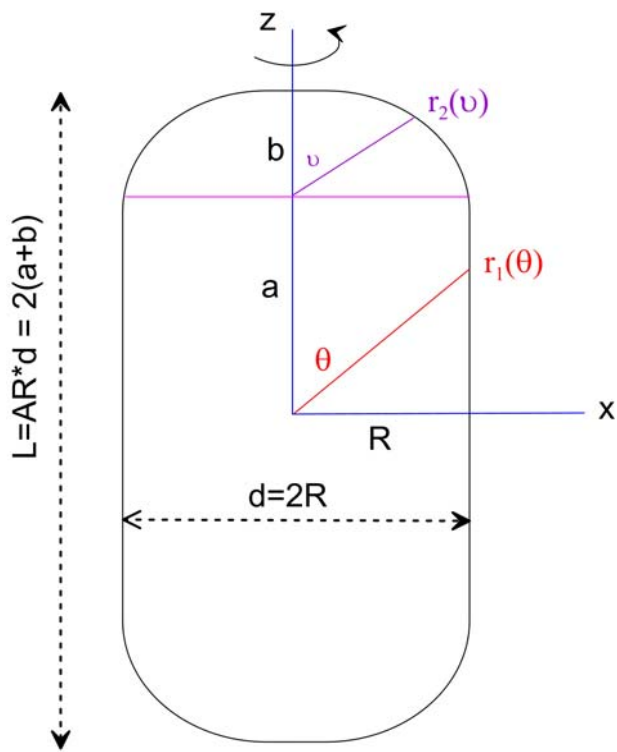

**Figure S20**. The geometry of a nanorod with the length $L$ L, diameter of $d$, and elliptical caps with the cap height $b = \chi_c R$ and radius $R = d/2$. The radius vectors $r_1(\theta)$ and $r_2(\vartheta)$ describe the boundary shape.

It is convenient to consider the cylindrical part of the nanorod surface, $S_1$, and the cap surface $S_2$. The total surface averaged enhancement factor is

S30

$$\langle EF \rangle = \frac{1}{S_1 + S_2}\left[\langle EF \rangle_1 + \langle EF \rangle_2\right] \tag{S13}$$

$$S_1 = 2\pi Ra = \pi R(L - 2b), \tag{S14}$$

$$S_2 = \pi Rb\left(q + \frac{\ln\left(q + \sqrt{q^2 - 1}\right)}{\sqrt{q^2 - 1}}\right), \tag{S15}$$

$$q = \frac{R}{b}. \tag{S16}$$

Equation (S15) gives a known expression for the surface area of an oblate spheroid with semiaxes $b$ and $R \geq b$. If $R = b$, then the right part of Eq. (S15) in brackets equals 1 and $S_2 = 2\pi R^2$ equals the semisphere surface. In the opposite limit $b \to 0$, the expression with logarithm equals zero and $S_2 = \pi R^2$, as it should be.

The surface averaged enhancement factor $\langle EF \rangle_1$ can be calculated by averaging over the polar angle $\theta$ or, equivalently, over the cylindrical coordinate $z$

$$\langle EF \rangle_1 = 2\pi R \int_0^a EF\left[r_1(z(\theta))\right] dz, \tag{S17}$$

$$r_1(z) = \sqrt{R^2 + z(\theta)^2}, \tag{S18}$$

$$z = R / \text{tg}(\theta), \tag{S19}$$

$$\text{tg}(\theta_{\min}) = \frac{R}{a}. \tag{S20}$$

Further, by using a general expression for any surface as given by rotation of an arbitrary curve $x = f(z)$ around the z-axis

$$S = 2\pi \int_0^b f(z)\sqrt{1 + [f'(z)]^2}\, dz, \tag{S21}$$

we get

$$\langle EF \rangle_2 = 2\pi Rb\sqrt{q^2 - 1}\int_0^1 \sqrt{p^2 + t^2}\, EF(r_2(t))\, dt, \tag{S22}$$

$$p^2 = \frac{b^2}{R^2 - b^2} = \frac{1}{(R/b)^2 - 1} = \frac{1}{q^2 - 1}, \tag{S23}$$

S31

$$r_2(t) = \sqrt{R^2 - t^2(R^2 - b^2)}, \tag{S24}$$

$$z = bt, \quad x = R\sqrt{1-t^2}, \quad r_2 = \sqrt{x^2 + z^2}. \tag{S25}$$

Equations (S25) relate the cap ellipse radius vector $r_2$ to the integration parameter $t$. If the enhancement factor in Eq. (S22) equals 1, then

$$\int_0^1 \sqrt{p^2 + t^2}\, dt = \left. \frac{t\sqrt{p^2+t^2}}{2} + \frac{p^2}{2}\ln\left|t + \sqrt{p^2+t^2}\right| \right|_0^1 = \frac{\sqrt{p^2+1}}{2} + \frac{p^2}{2}\ln\frac{1+\sqrt{p^2+1}}{p}, \tag{S26}$$

and we arrive at Eq. (S15) for the cap surface area.

The numerical evaluation of integrals (S17) and (S22) was carried out by using COMSOL codes:

Int_Snp(withsol('sol1',(up(ewfd.normE))^2/E0^2,setval(lambda0,wlL),setval(p,p))*withsol('sol1',(up(ewfd.normE))^2/E0^2,setval(lambda0,wlR),setval(p,p)))/Int_Snp(1).

Here, Int_Snp(1) is an operator of integration over the particle surface (it can be split by the sum of two integrals over the nanorod end and the cylindrical part; see Eq. S13); wlL and wlR are the laser and Raman wavelength, respectively. As we have two Raman lines, we used two analogous syntaxic procedures for wlR1 and wlR2). Finally, the expression ewfd.normE^2/E0^2 means the normalized squared modulus of the local field; the operator "up" points to the outer side. For details concerning other operators, the readers are referred to Refs.[5-7]

For orientation averaging, we used the following COMSOL options: "Data Series Operation —> Average" and "Method → Integration."

**Section S2.8. Morphological parameters of AuNSTs.**

**Table S4**. The average morphological parameters and LPR wavelengths of etched nanostars.

| Sample | 1 | 2 | 3 | 4 | 5 | 6 | 7 | 8 |
|---|---|---|---|---|---|---|---|---|
| LPR (nm) | 838 | 815 | 764 | 697 | 660 | 626 | 598 | 565 |
| Tip length (nm) | 25±4 | 22±4.8 | 21±4.5 | 16±4.5 | 13±4.6 | 12±4.3 | 13±3 | 9.9±2.7 |
| Tip angle (degrees) | 27±9.3 | 27±9.8 | 32±10 | 38±11 | 44±13 | 54±16 | 61±17 | 75±20 |
| Core diameter (nm) | 30±3.6 | 30±3.6 | 30±3.2 | 30±2.8 | 31±2.9 | 33±3.7 | 36±4.0 | 36±3.5 |

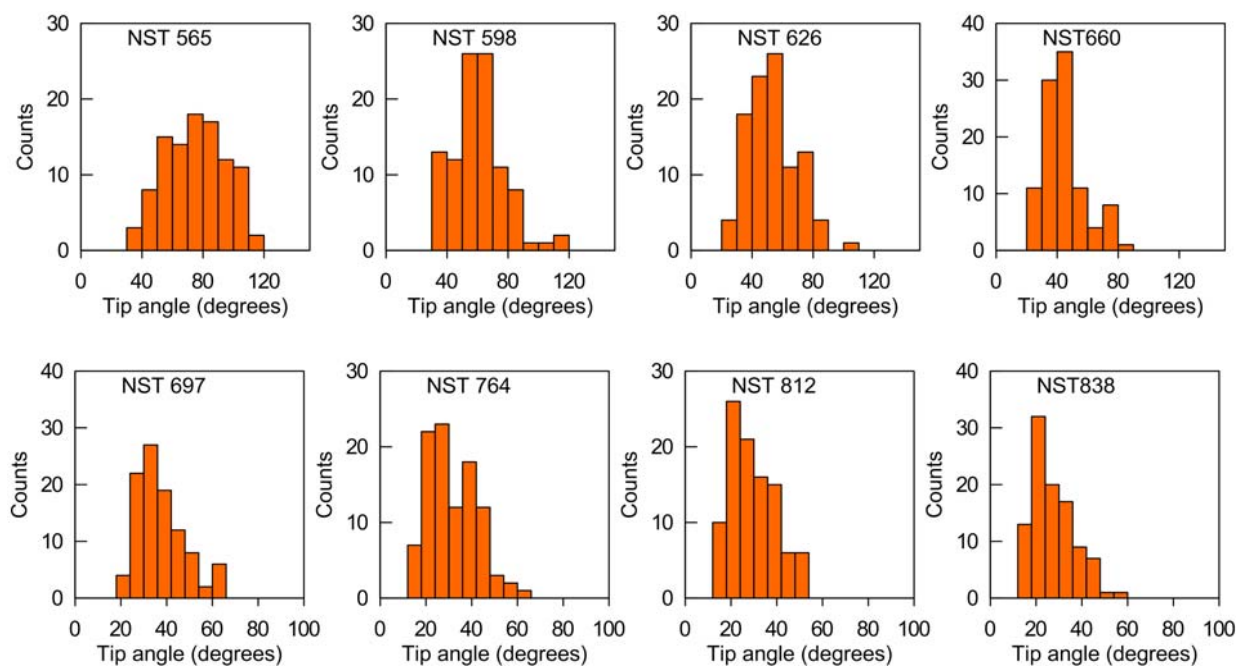

**Figure S21**. Histograms of the tip angle distributions for 8 samples NST 565-838.

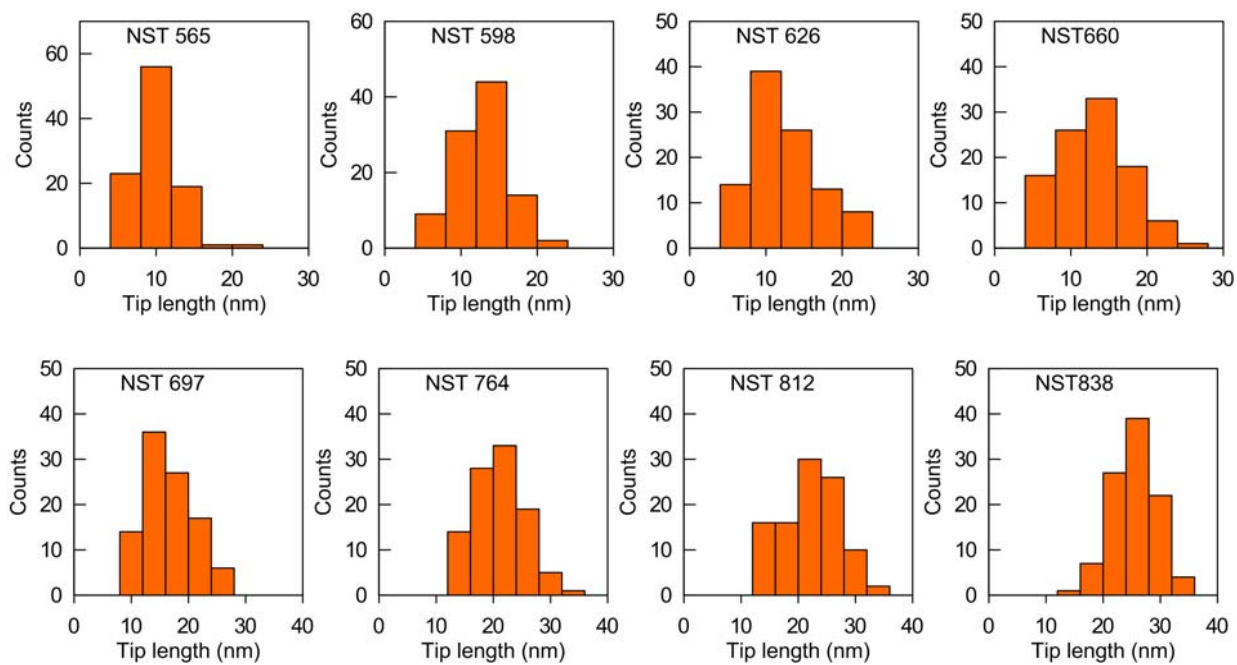

**Figure S22**. Histograms of the tip length distributions for 8 samples NST 565-838.

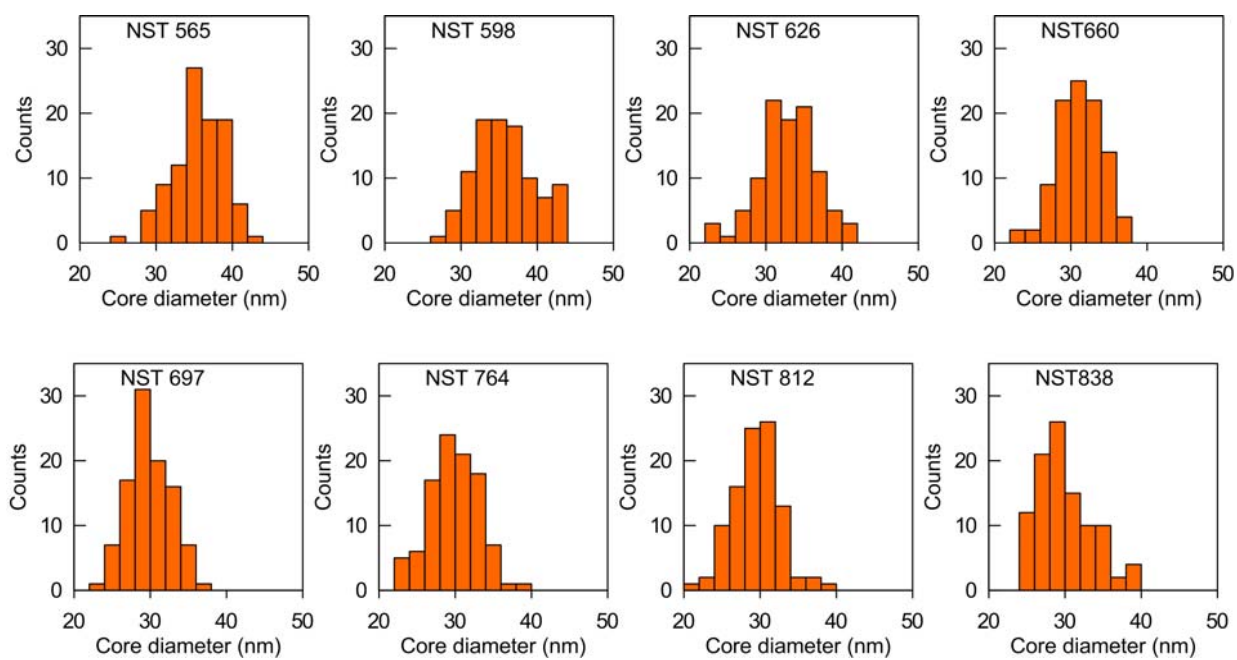

**Figure S23**. Histograms of the core diameter distributions for 8 samples NST 565-838.

**Section S2.9. SERS spectra for AuNST@NBT and AuNST@Cy7.5 samples at 785 nm and 633-nm laser excitation**

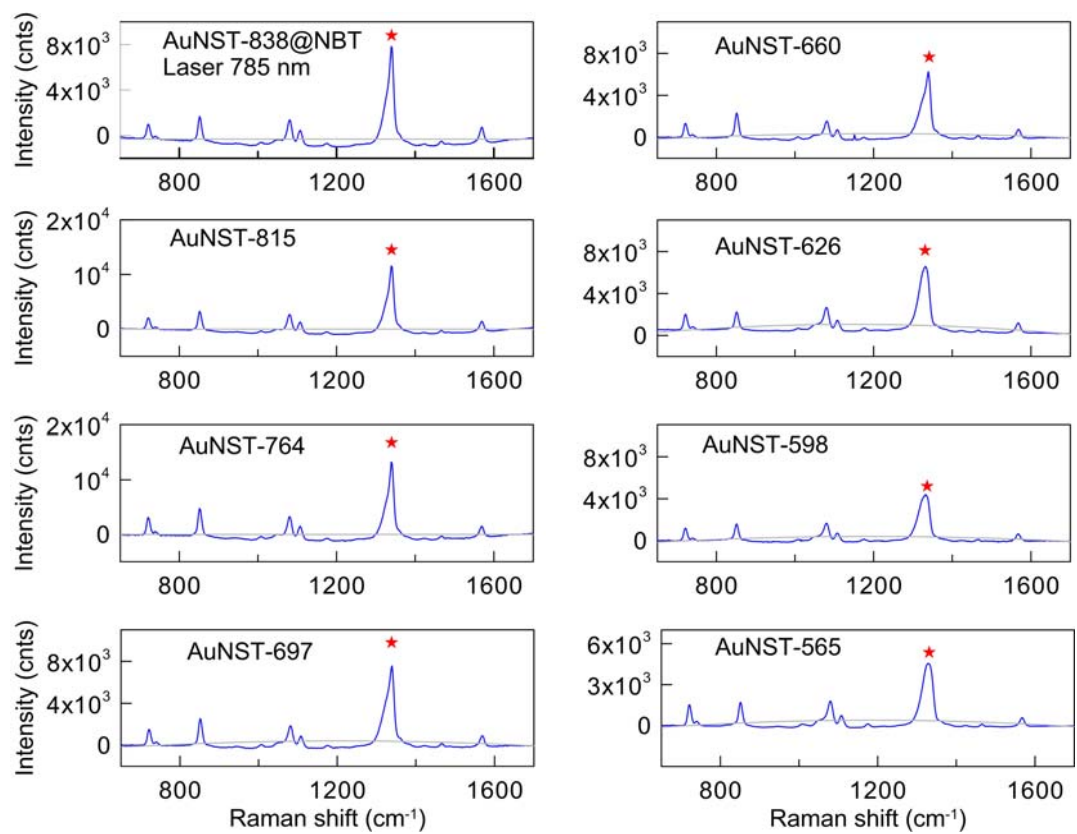

**Figure S24**. SERS spectra for AuNST@NBT samples measured at 785 nm laser excitation. The LPR wavelength decreases from 850 to 570 nm. The average fitted baseline (gray) was subtracted from the original spectra, thus explaining the negative values under the baseline.

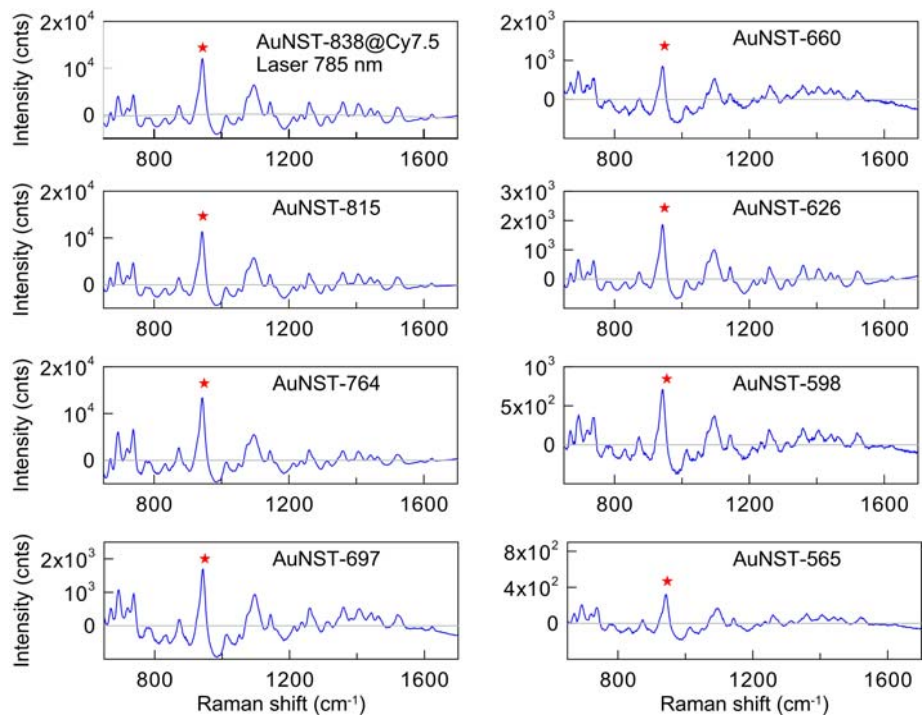

**Figure S25**. SERS spectra for AuNST@Cy7.5 samples measured at 785 nm laser excitation. The LPR wavelength decreases from 850 to 570 nm. The average fitted baseline (gray) was subtracted from the original spectra, thus explaining the negative values under the baseline.

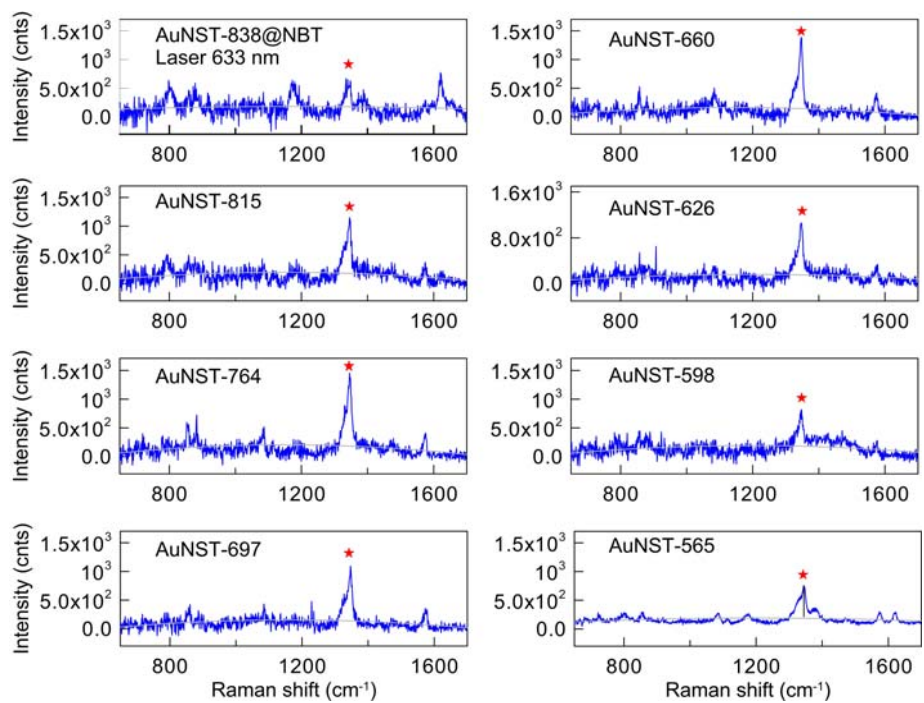

**Figure S26**. SERS spectra for AuNST@NBT samples measured at 633 nm laser excitation. The LPR wavelength decreases from 838 to 570 nm. The average fitted baseline (gray) was subtracted from the original spectra, thus explaining the negative values under the baseline.